\documentclass[preprint]{aastex631}

\usepackage{braket}

\begin{document}

\title{Correlations and Distinguishability Challenges in Supernova Models: Insights from Future Neutrino Detectors}

\author[0000-0001-6760-1028]{Maria Manuela Saez}
\affiliation{RIKEN Interdisciplinary Theoretical and Mathematical Sciences Program (iTHEMS), 2-1 Hirosawa, Wako, Saitama 351-0198, Japan}
\affiliation{Department of Physics, University of California, Berkeley, CA 94720}
\author[0000-0002-3222-7010]{Ermal Rrapaj}
\affiliation{RIKEN Interdisciplinary Theoretical and Mathematical Sciences Program (iTHEMS), 2-1 Hirosawa, Wako, Saitama 351-0198, Japan}
\affiliation{Lawrence Berkeley National Laboratory, One Cyclotron RD, Berkeley, CA, 94720, USA}
\affiliation{Department of Physics, University of California, Berkeley, CA 94720, USA}

\author[0000-0003-1409-0695]{Akira Harada}
\affiliation{RIKEN Interdisciplinary Theoretical and Mathematical Sciences Program (iTHEMS), 2-1 Hirosawa, Wako, Saitama 351-0198, Japan}

\author[0000-0002-7025-284X]{Shigehiro Nagataki}
\affiliation{Astrophysical Big Bang Laboratory (ABBL), RIKEN Cluster for Pioneering Research, 2-1 Hirosawa, Wako, Saitama 351-0198, Japan}
\affiliation{RIKEN Interdisciplinary Theoretical and Mathematical Sciences Program (iTHEMS), 2-1 Hirosawa, Wako, Saitama 351-0198, Japan}
\affiliation{Astrophysical Big Bang Group (ABBG), Okinawa Institute of Science and Technology Graduate University, 1919-1 Tancha, Onna-son,Kunigami-gun, Okinawa 904-0495, Japan}

\author[0000-0002-3146-2668]{Yong-Zhong Qian}
\affiliation{School of Physics and Astronomy, University of Minnesota, Minneapolis, Minnesota 55455, U.S.A.}



\begin{abstract}

This paper explores core-collapse supernovae as crucial targets for neutrino telescopes, addressing uncertainties in their simulation results. We comprehensively analyze eighteen modern simulations and discriminate among supernova models using realistic detectors and interactions. A significant correlation between the total neutrino energy and cumulative counts, driven by massive lepton neutrinos and oscillations, is identified, particularly noticeable with the DUNE detector. Bayesian techniques indicate strong potential for model differentiation during a Galactic supernova event, with HK excelling in distinguishing models based on equation of state, progenitor mass, and mixing scheme.
\end{abstract}

\keywords{Core-collapse supernovae (304) --- Supernova neutrinos (1666) --- Neutrino telescopes (1105) --- Bayesian statistics (1900)}


\section{Introduction}

Core-collapse supernovae (SN) mark the final stage of stellar evolution for stars with masses exceeding $8M_\odot$, and they stand as awaited targets for neutrino telescopes.  A mere 1\% of the gravitational binding energy is transformed into kinetic energy during compact object formation, while the remaining 99\% is transported by neutrinos of various flavors, each with energies in the range of several MeV \citep{Woosley:2002}. The primary mechanisms responsible for neutrino production within the SN core encompass electron capture by nucleons ($e^{-}+p\rightarrow n+\nu_e$), pair annihilation ($e^+ + e^- \rightarrow \nu_e+\bar{\nu_e}$), and nucleon Bremsstrahlung ($N+N\rightarrow N'+N'+\nu+\bar{\nu}$) \citep{Janka:2017}. After traversing stellar matter and space, neutrinos that eventually reach Earth become valuable tools for unraveling the secrets of the stellar core.

By studying the unique signatures that neutrinos imprint on detectors, with effective discrimination among neutrino flavors, we gain insight into their underlying physics. Presently, a number of detectors eagerly await the detection of SN neutrinos from the next galactic explosion, while three significant future detectors are either under construction or in the planning stages: Hyper-Kamiokande (HK), The Deep Underground Neutrino Experiment (DUNE), and The Jiangmen Underground Neutrino Observatory (JUNO). These detectors intercept SN neutrinos through their electroweakly or strongly interacting products, involving weak charged-current (CC) and neutral-current (NC) interactions with electrons and nuclei. Among the pertinent interaction channels for both current and future detectors are inverse beta decay and neutrino-proton elastic scattering for scintillation detectors, inverse beta decay and neutrino-electron elastic scattering for water Cherenkov detectors, and absorption interactions with $^{40}$Ar in liquid Argon time projection chambers, among others.

Although the core-collapse supernova explosion mechanism is partially understood through computer simulations and observations of SN1987A's neutrino burst, many details remain uncertain. Simulations using different approximations due to computational limitations yield varied results, both quantitatively and qualitatively, even for identical progenitors \citep{OConnor:2018}. Therefore, it is crucial to seize the upcoming galactic supernova event as a unique opportunity to compare model predictions with observations and determine the most accurate representation of reality.
Due to the scarcity of observed supernova neutrinos, current research heavily relies on computer simulations. Different research groups worldwide conduct simulations of supernovae using progenitors with diverse masses, compositions, rotational velocities, and other characteristics. These simulations vary in complexity, ranging from highly detailed models that require substantial computational resources to simulate individual or very few progenitors accurately, to simplified studies that impose rotational or spherical symmetry and employ approximations in neutrino transport and other physical parameters to reduce computational demands. While the latter simulations may be less realistic, they enable researchers to conduct a large number of simulations and investigate the impact of varying progenitor parameters (such as mass or metallicity), numerical resolution, and specific approximations.

Various current and next-generation neutrino detection collaborations have assessed the capabilities of their detectors in observing supernova neutrinos \citep{Wang:2021, Abi:2021,  Abe:2018}. These collaborations typically utilize a limited number of supernova models as benchmarks to demonstrate the expected event counts in their detectors, their sensitivity to different interaction channels, and their ability to distinguish supernova neutrinos from background sources. Depending on the detector's sensitivity, these collaborations may also conduct more advanced analyses, such as detecting pre-supernova neutrinos or determining the direction of a supernova \citep{Abe:2016, KamLAND:2015}.

Conducting a study that takes into account both detector efficiencies, reconstruction uncertainties, and background considerations, and moreover, doesn't solely focus on a limited set of benchmark models, but rather conducts a more generalized study, is a challenging task. However, it is necessary to comprehensively explore the diverse features observed in modern computer simulations. 

Despite the extensive ongoing research in this field, not too much analysis has been conducted to date that demonstrates the level of discrimination achievable between different computer models of supernovae using several realistic detectors and interaction channels. Notable contributions in this direction include works by \cite{Loredo:2002}, \cite{Olsen:2022},  \cite{Abe:2021}, and \cite{Damiano:2023}. 

In this study, our goal is to delve into Supernova Models, examining correlations and assessing the power of distinguishability within the framework of future neutrino detectors. As for the former, we have detected a strong correlation between the cumulative number of events  and the total energy emitted by neutrinos. This correlation remains independent of the considered progenitor mass, the dimensionality of the simulation, the equation of state used, and the effects of stellar rotation.
Regarding the latter, we employ Bayesian statistics to test the distinguishability of pairs of SN models and study the feasibility of distinguishing between different equations of states, progenitor masses, or neutrino oscillation schemes. We find that all models can be distinguished from each other by a detector such as Hyper-Kamiokande.

The paper is organized as follows. In Section \ref{sec:models}, we describe the sample of SN models studied. In sections \ref{sec:fluxes} and \ref{sec:nu_signal}, we describe the details in the calculation of the neutrino fluxes, the incorporation of the oscillations, and the calculation of the expected signal in the three future detectors considered. In section \ref{sec:result_nu_signal}, we present the results of the expected signals and the found correlations. In section \ref{sec:model_discr}, we address the problem of distinguishability between models through a Bayesian statistical analysis. Finally, in section \ref{sec:conclusions}, we present the conclusions.

\section{SN models} \label{sec:models}
In this paper, we utilize two series of supernova simulations, 3D Fornax and 2D Boltzmann-radiation-hydrodynamics models. Supernova simulation codes are characterized by dimensionality, neutrino transport, neutrino reactions, and gravity\footnote{Magnetic fields may be another aspect, but all models considered here are non-magnetic.}. The 3D and 2D models adopted in this paper differ in almost all aspects. However, both models treat three neutrino species, i.e., the electron-type neutrinos $\nu_{\rm e}$, the electron-type antineutrinos $\bar{\nu}_{\rm e}$, and the heavy-lepton-type neutrinos $\nu_x$. Here, $\nu_x$ is the representative of the other four species of neutrinos, $\nu_\mu$, $\bar{\nu}_\mu$, $\nu_\tau$, and $\bar{\nu}_\tau$, because they has the same reaction rates under the employed neutrino reaction sets and hence has the identical luminosities and spectra. The possible modification for $\nu_x$ radiation is the flavor conversion discussed in section \ref{sec:fluxes}: in this case, we need to split $\nu_x = \nu_\mu,\,\nu_\tau,\,\bar{\nu}_\mu,\,\bar{\nu}_\tau$ into $\nu_x = \nu_\mu,\,\nu_\tau$ and $\bar{\nu}_x = \bar{\nu}_\mu,\,\bar{\nu}_\tau$. These two notations can be easily distinguished from the context. The detailed description of the code is presented in the following.

\subsection{3D Fornax models}
The \emph{3D models} employed in this paper are based on the core-collapse supernova (CCSN) neutrino estimate \citep{Burrows:2019b,Vartanyan:2018} obtained from the 3D FORNAX code. They solve the 3D neutrino-radiation hydrodynamics with self gravity in a self-consistent manner (see \cite{Skinner:2019} for more details). They employed the M1-closure scheme with $12$ energy groups for neutrino transport. The neutrino reactions are electron captures on proton and nuclei, anti-electron capture on neutron, nucleon scattering, coherent scattering off nuclei, pair production, nuclear bremsstrahlung, and electron scattering. 
For gravity, they adopted the treatment of Marek case A \citep{Marek:2006}, where the monopole component of the Newtonian gravitational potential is replaced by the solution of the Tolman-Oppenheimer-Volkoff (TOV) equation. Furthermore, they imposed the velocity perturbation of $\sim 1\%$ at $10\,{\rm ms}$ after the core bounce to facilitate the prompt convection.

The 3D simulations are accessible at \url{https://www.astro.princeton.edu/~burrows/nu-emissions.3d/} and encompass progenitors with masses ranging from 9 to 60 M$_\odot$\footnote{All the Fornax simulations in this set were based on the initial models described in \cite{Sukhbold:2016}, with the exception of the 25 M$_{\odot}$ model. The latter incorporates notable variations in the model implementation, including adjustments to surface boundary pressure, pair neutrino losses, and a significantly finer mass resolution. For more information, please refer to the details provided in \cite{Sukhbold:2018}.}.
 They employed the SFHo equation of state (EoS) \citep{Steiner:2013} based on the relativistic mean field (RMF) theory whose parameters are tuned to fit the observations. The nuclear statistical equilibrium (NSE) is considered for the ensemble of nuclei in order to calculate the thermodynamical and statistical properties of nonuniform matter.
For further details regarding these 3D models, please refer to \citep{Vartanyan:2018, Burrows:2019, Burrows:2019b}.

A related study presented in Ref. \citep{Nagakura:2021} investigated the expected neutrino signals from these models in representative underground neutrino detectors. 

\subsection{2D Boltzmann-radiation-hydrodynamics models}
The \emph{2D models} employed in this paper are taken from the two-dimensional axisymmetric simulations of the Boltzmann-radiation-hydrodynamics code \citep{Sumiyoshi:2012, Nagakura:2014, Nagakura:2017, Nagakura:2019}. They solve the Boltzmann equation on phase space for neutrino transport coupled with the hydrodynamics and Newtonian gravity. For computational cost reasons, they adopt the results of 2D simulations. The neutrino reaction set is similar to that of 3D models, but details are slightly different (e.g., electron capture on light nuclei in some of 2D models, many-body correction on nucleon scattering in 3D models, and so on). The Newtonian gravitational potential is obtained by solving the Poisson equation. The 2D models imposed the $0.1\%$ velocity perturbation at $< 1\,{\rm ms}$ after the core bounce for the prompt convection.

The simulations considered in this work are based in three different EoS: Furusawa and Shen (FS), Lattimer and Swesty (LS), and  Furusawa and Togashi  (FT). The FS EoS is based on the Shen EoS \citep{Shen:1998}, and the NSE composition model. 
The Shen EoS, the basis of the FS EoS, models the strong interaction by the RMF theory with the TM1 parameter set. On the other hand, the LS EoS is based on the liquid drop model of the nuclei and the Skyrme-type interaction. As for the composition, Lattimer and Swesty \citep{Lattimer:1991} assume that the heavy nuclei are represented by a single nuclear species (single nuclear approximation, SNA), and only the alpha particle is considered as the light nuclei. Also, we studied the FT EoS is based on Togashi variational method but NSE is assumed \citep{Togashi:2017}.  We have conducted a comprehensive study of the aforementioned equations of state for progenitor masses of 11.2, 15, and 27 solar masses. We also consider the rotational stellar core collapse simulations for some models. The 2D models are named by the EoS, the progenitor mass, and the rotation: for example, ``FS 11.2\,M$_\odot$'' is the non-rotating model with the FS EoS and the ZAMS mass of 11.2\,M$_\odot$, while ``FS 11.2\,M$_\odot$ rot'' is the rotational counterpart. The details of the models are presented in \cite{Nagakura:2018, Harada:2020} (FS 11.2\,M$_\odot$, LS 11.2\,M$_\odot$), \cite{Iwakami:2022} (LS 15\,M$_\odot$), \cite{Harada:2019} (FS 11.2\,M$_\odot$ rot), \cite{Nagakura:2019b} (FT 11.2\,M$_\odot$), and the forthcoming papers (LS 27\,M$_\odot$, FT 15\,M$_\odot$, FT 27\,M$_\odot$, and FT 15\,M$_\odot$ rot). Notably, this study represents the first instance where predictions have been made regarding the neutrino signal obtained using the Boltzmann-radiation-hydrodynamics code.

\section{Flux calculations}\label{sec:fluxes}

The neutrino flux can be expressed as (for the un-oscillated case): 
\begin{equation}
    \frac{d\phi_{\nu_\alpha} (E_\nu,t)}{dE_\nu}=\frac{L_{\nu_\alpha} (t)}{4\pi d^2\braket{E_{\nu_\alpha} (t)}}f_{\nu_{\alpha}}(E_\nu,t) \, , \label{flux}
\end{equation}
where $f_{\nu_{\alpha}}(E_\nu,t)$ represents the normalized distribution, which is commonly described using a two-parameter fit \citep{Keil:2003}. This fit allows for deviations from a strictly thermal spectrum, often chosen due to its analytical simplicity and given by 
\begin{equation}
    f_{\nu_{\alpha}}(E_\nu,t)=\mathcal{N} \left(\frac{E_\nu}{\braket{E_{\nu_\alpha} (t)}}\right)^{\beta_{\nu_\alpha} (t)} \exp\left[{-(\beta_{\nu_\alpha}(t)+1)}\frac{E_\nu}{\braket{E_{\nu_\alpha} (t)}} \right] \label{distro-espectral}
\end{equation}
where $\mathcal{N}$ is the normalization,  $\alpha$ stands for an specific lepton flavor, $\beta_{\nu_\alpha}$ is the pinching parameter of the spectral distribution, and $\braket{E_{\nu_\alpha}}$ are the neutrino mean energies. The pinching parameter is related to the neutrino mean energy and mean-square energy $\braket{E_{\nu_\alpha}^2}$ by
\begin{equation}
    \frac{\braket{E_{\nu_\alpha}^2}}{\braket{E_{\nu_\alpha}}^2} = \frac{2+\beta_{\nu_\alpha}}{1+\beta_{\nu_\alpha}}. \label{pinching}
\end{equation}
The 3D models provide the time evolution of $\braket{E_{\nu_\alpha}}$ and $\braket{E_{\nu_\alpha}^2}$, and hence the pinching parameter is calculated from these quantities. The 2D models provide the full spectral information, but we calculated $\braket{E_{\nu_\alpha}}$ and $\braket{E_{\nu_\alpha}^2}$ from the spectra and determine the pinching parameter through the above relation instead of directly fitting equation (\ref{distro-espectral}).

Neutrino flavor conversions inside a supernova are possible and their effectiveness depends on the matter density. Since the matter density (and therefore the potencial) decreases with the star radius, active neutrinos exhibit two Mikheyev-Smirnov-Wolfenstein (MSW) resonances called H (high density) and L (low density) where the flavor conversion mechanism is amplified \citep{Dighe:1999}. 
The neutrino flux $F_{\nu_\alpha}$ that arrives at Earth can be computed as a superposition of the neutrino fluxes $F_{\nu_\alpha}^0$ produced in the neutrinosphere inside the SN. 
The flux for
$\nu_e$ and $\nu_x$ neutrinos (anti-neutrinos) is written:
\begin{eqnarray}
F_{\nu_e}&=& P_{e} F^0_{\nu_e} + (1-P_e)F^0_{\nu_x} \, \, \, , \nonumber \\
F_{\bar{\nu}_e}&=& \bar{P}_e F^0_{\bar{\nu}_e} + (1-\bar{P}_e)F^0_{\bar{\nu}_x} \, \, \, , \nonumber \\
F_{\nu_x}&=& \frac{1}{2}(1-P_e) F^0_{\nu_e} + \frac{1}{2}(1+P_e)F^0_{\nu_x}\, \, \, , \nonumber \\
\label{flujos-3}
F_{\bar{\nu}_x}&=& \frac{1}{2}(1-\bar{P}_e) F^0_{\bar{\nu}_e} + \frac{1}{2}(1+\bar{P}_e)F^0_{\bar{\nu}_x}\, \, \, .
\end{eqnarray}
 In the previous equations $P_e$ and $\bar{P}_e$ are the survival probabilities for $\nu_e$ and ${\bar{\nu}_e}$ respectively, which depend on the neutrino-mass hierarchy. That is:
\begin{eqnarray}
P_e&=&|U_{e1}|^2 P_H P_L +|U_{e2}|^2 P_H (1-P_L) + |U_{e3}|^2(1-P_H) \, \, \, , \nonumber \\
\bar{P}_e&=& |U_{e1}|^2 \, \, \, ,
\end{eqnarray}
for normal mass ordering (NMO), and
\begin{eqnarray}
P_e&=&|U_{e1}|^2P_L+|U_{e2}|^2(1-P_L)  \, \, \, , \nonumber \\
\bar{P}_e&=& |U_{e1}|^2\bar{P}_H+|U_{e3}|^2(1-\bar{P}_H) \, \, \, ,
\end{eqnarray}
for inverse mass ordering (IMO), respectively.

In last equations $U_{ek}$ $(k=1,\, 2,\, 3)$ are the components of the first row of the PMNS-matrix.  
$P_H$ ($\bar{P}_H$) and $P_L$ ($\bar{P}_L$) are the neutrino (antineutrino) crossing probabilities for the H and L resonances respectively. In the case of adiabatic transformations, the probabilities are reduced to $P_e=|U_{e3}|^2$ and $\bar{P}_e=|U_{e1}|^2$ for NMO, and $P_e=|U_{e2}|^2$ and $\bar{P}_e=|U_{e3}|^2$ for IMO.

\subsection{Spectral parameters for the SN models}\label{spectral_info}

\begin{figure}[h]
 \centering
        \includegraphics[width=1.\textwidth]{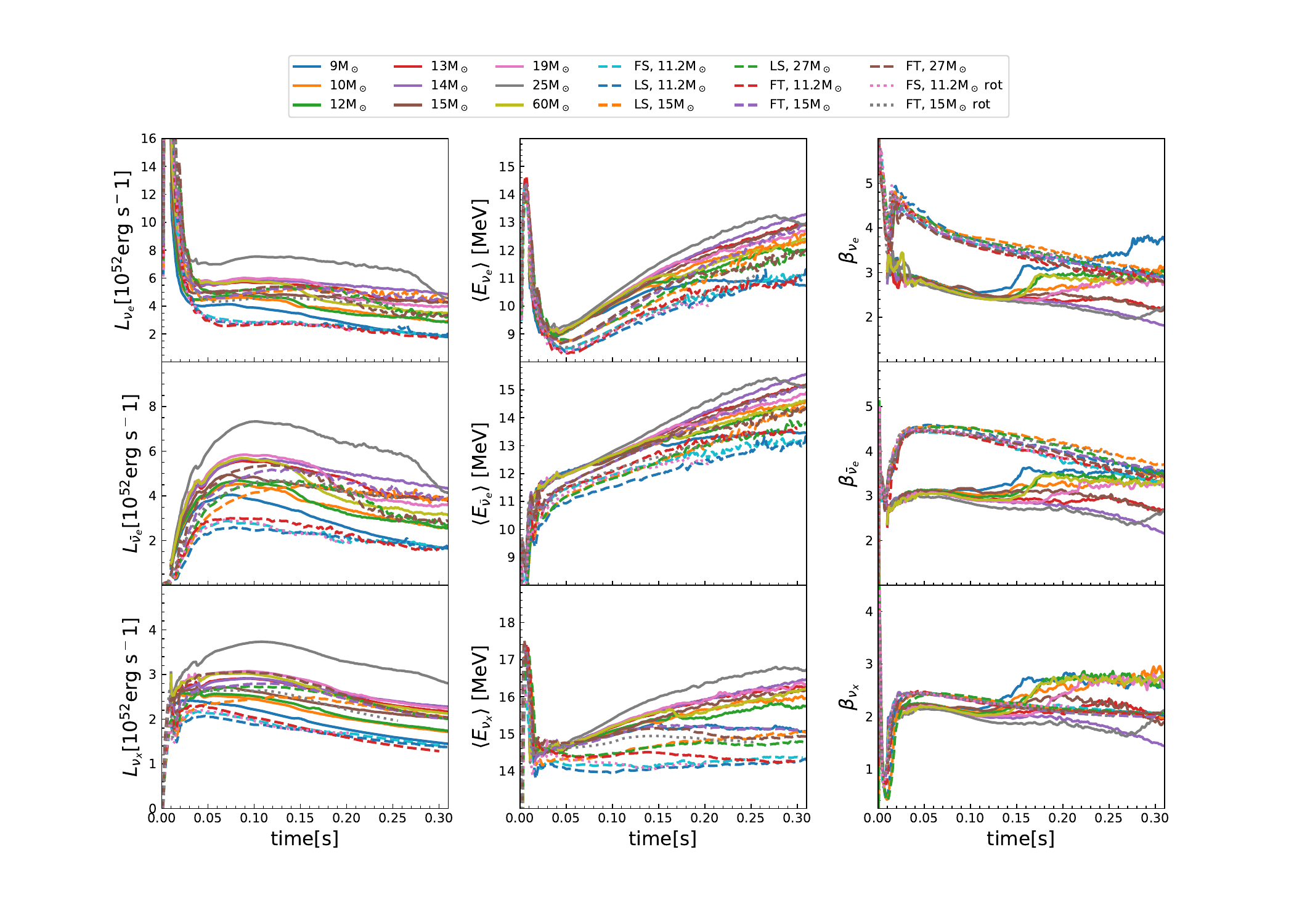} \caption{Left panel: Angle-integrated luminosities as a function of time for the different progenitor masses and EoSs. Middle panel: Average angle-integrated energy as a function of time for the different progenitor masses and EoSs. Right panel: Spectral shape parameter (pinching) $\beta$ as a function of time for the different progenitor masses and EoSs. Solid and dashed lines refer to 3D simmulations and 2D axi-summetric simmulations respectively. 2D rotational simmulations are shown in dotted lines. Upper, middle, bottom rows represent  $\nu_e$, $\bar{\nu}_e$, $\nu_x$, respectively. 
        }\label{spectral_param_vs_t}
    \end{figure}

The behavior of the mean energies $\braket{E_{\nu_\alpha} (t)}$, the spectral shape parameters (pinching) $\beta_{\nu_\alpha}(t)$  and luminosities $L_{\nu_\alpha}(t)$ as a function of the time  are shown in the figure \ref{spectral_param_vs_t} for all the models considered. Our analysis is limited to 300\,ms, as it corresponds to the longest time reached by all the simulations.

The luminosities and average neutrino energies are comparable between the 3D and 2D models.  Although there is a trend that higher progenitor masses correspond to greater luminosity, this relation is not strictly monotonic. Specifically, the 25 M$_{\odot}$ model emerges as the brightest. This particular model achieves the highest luminosities due to its elevated compactness parameter (see \cite{Burrows:2019b}). The compactness parameter serves as a metric gauging the shallowness of the mass density profile within the inner progenitor. It significantly impacts the evolution of the infall accretion rate, consequently affecting both neutrino luminosities and neutrino energies.
In cases where the progenitor mass and equation of state remain the same, rotational models exhibit lower luminosities and neutrino spectra. Regarding the EoS dependence, we observe that $L_{\nu_e}(FT) < L_{\nu_e}(LS) < L_{\nu_e}(FS)$, and $L_{\bar{\nu}_e}(LS) < L_{\bar{\nu}_e}(FS) < L_{\bar{\nu}_e}(FT)$.
Given the parametric expression (\ref{distro-espectral}) for neutrino fluxes, we can see that greater values of the average energies shifts the maximum of the distribution to the high-energy side. 
In general, the higher the $\braket{E}$, the counts in the detector increase. 

On the other hand, smaller values of $\beta$ increase the amount of neutrinos at higher energies and reduce the number of neutrinos at lower energies (see equation (\ref{pinching})). With regard to this parameter we notice a separation between the 3D and 2D models. Specifically, the values of $\beta$ for $\nu_e$ and $\bar{\nu}_e$ are lower in the 3D models than the 2D ones for the time range up to 0.2 seconds. For the 2D models it peaks before 0.05 seconds and then proceeds to decrease. The same pattern can be seen for $\nu_x$ but to a lesser extent. As such, we expect the 2D models to have a higher count rate of neutrinos at lower energies.

\section{Expected signal in underground detectors}\label{sec:nu_signal}
To calculate the expected number of events per time bin for ideal detectors we consider:
\begin{equation}
    \frac{dN(t)}{dt}=N_{tar}\int \int \frac{dF_{\nu}}{dE_\nu}(E_\nu,t) \frac{d\sigma}{dE'} (E_\nu, E') dE' dE_\nu
\end{equation}
where $N_{tar}$ is the number of targets within the fiducial volume, $E_\nu$ is the incoming neutrino energy, $E^\prime$ is the energy of an outgoing particle depending on the detection channel, $\frac{dF_{\nu}}{dE_\nu}$ is the spectral neutrino flux at the detector, and $\frac{d\sigma}{dE^\prime} (E_\nu,E^\prime)$ is the energy-differential cross section for the reaction. No background noise is assumed in this expression.

\begin{figure}[h]
  \centering
        \includegraphics[width=.6\textwidth]{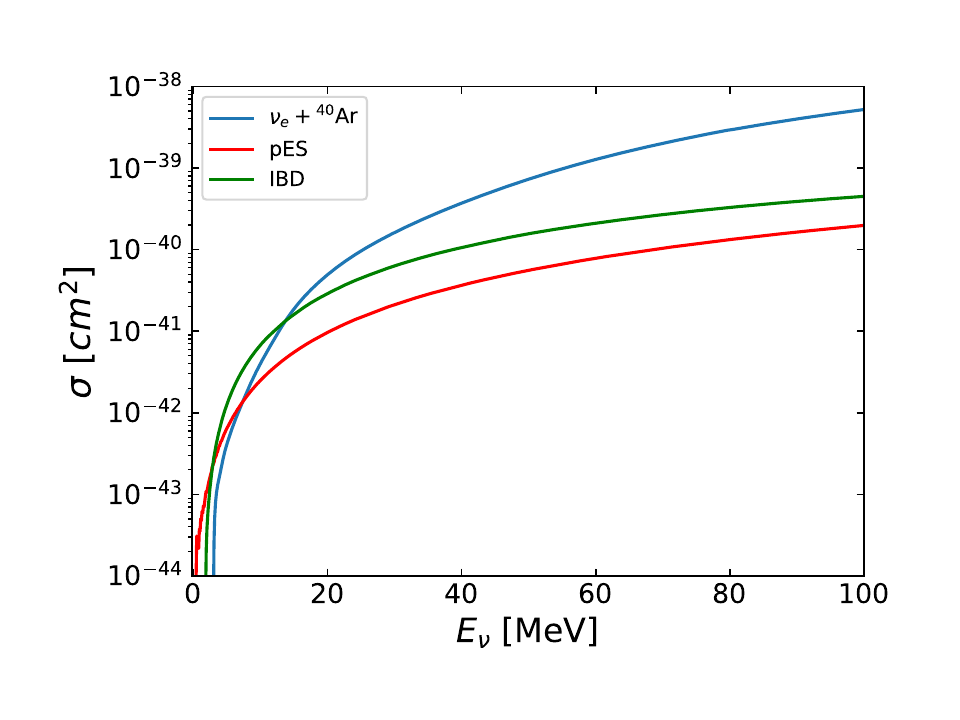} \caption{The total cross sections are plotted as a function of neutrino energy for the interaction processes studied in this work.}\label{xs}
    \end{figure}

To be more realistic, we calculate the expected signal by accounting for both detector efficiency and smearing effects,

\begin{equation}
        \frac{dN(t)}{dt}=N_{tar}\int_{E_{th_{d}}}^{E_{max}} dE \int^{\infty}_{E_{th_{r}}} dE_\nu \int_0^\infty dE^\prime \epsilon(E')\frac{dF_{\nu}}{dE_\nu}(E_\nu,t) \frac{d\sigma}{dE'} (E_\nu, E') G(E,E',\delta), 
\end{equation}

where E is the visible observed energy in the detector and ${E_{th_{d}}}$ and $E_{max}$ its thresholds. $\epsilon(E')$ is the detection efficiency. Because of smearing, an outgoing particle of energy $E'$ may be detected at energy E, this effect can be appoximated including a Gaussian function $G(E,E',\delta)$ for the energy resolution with the central value $E'$ and the standard width $\delta$. $E_{th_{r}}$ is the threshold for the reaction. 
In both cases, to obtain the expected number of events, we integrate over time (up to 0.3\,s).  
 The interaction channels considered in this work correspond to inverse beta decay (IBD), neutrino proton elactic scattering (pES) and electron neutrino charged current absorption on Argon ($\nu_e\, + $ Ar). The cross sections for the mentiones process are shown in Figure \ref{xs}. We expect the typical neutrino energy to in the 10-20 MeV range for the initial time of the core collapse, and thus, the typical cross sections are in the range $10^{-42}-10^{-40}$ cm$^{2}$. In the next sections we describe in detail the detectors and respective cross sections plotted in this figure.

\subsection{Hyper-Kamiokande}

HK is a next-generation water Cherenkov detector that will be built near the  currently operating Super-Kamiokande detector \citep{Abe:2018} .
The design of HK is similar to that of Super-Kamiokande. It is a large, cylindrical detector with a height of 71 m and diameter of 68 m, filled with  ultra-pure water.  The inner detector will provide a fiducial volume of 188 kt \citep{Yano:2022}.

The dominant interaction will be the IBD $\left(\bar{\nu_e} + p \rightarrow n + e^+\right)$. Because the recoil of the proton is negligible, the differential-cross section for the IBD is $\frac{d\sigma}{dE^\prime}(E_\nu,E^\prime) = \sigma_{\bar{\nu}_{\rm e}}(E_\nu) \delta(E^\prime - E_\nu + \Delta)$, where $\delta$ in this expression is the delta function and $\sigma_{\bar{\nu}_{\rm e}}(E_\nu)$ is the total cross section approximated as  
\begin{equation}\label{sig(e)}
\sigma_{\bar{\nu}_{e}}(E_\nu)=10^{-43} \rm{cm}^2p^\prime E^\prime E_\nu^{-\,0.07056\, +\,0.02018\, y\,-\,0.001953 \, y^3} \, \, ,
\end{equation}
for energies lower than $300 \, {\rm MeV}$ \citep{Strumia:2003}. In this equation, all variables with the dimension of energy are measured in units of MeV. The symbol $p^\prime$ stands for the positron momentum related to the neutrino energy  $E_\nu$  as  $p^{\prime 2} = (E_\nu - \Delta)^2 - m_{\rm e}^2$, where the neutron to proton mass difference is $\Delta=m_n-m_p=1.293 \, {\rm MeV}$ and $m_e$ is the positron mass. The neutrino and positron energy are related by  $E^\prime = E_\nu - \Delta$ owing to the delta function, and $y= \ln E_\nu $ in the exponent of Eq. (\ref{sig(e)}). For this channel $E_{{th}_r}=1.806 \, \rm{MeV}$ is the threshold energy.

For the detector configuration we consider a fiducial volume of 188 kt with a number of free protons $N_p=1.2567\times 10^{34}$,  and a smearing function modeled with a resolution 
$\delta/\textrm{MeV}= -0.0839 + 0.349 \sqrt{E^\prime/\textrm{MeV}} + 0.0397 E^\prime/\textrm{MeV}$.
The exact photosensor configuration of Hyper-Kamiokande has not yet been determined, so for our calculations, we consider a perfect efficiency.  We set the threshold $E_{{th}_d}=7$ MeV and the cut-off energy $E_{max}=100$ MeV \citep{Abe:2018}. 

\subsection{JUNO}
JUNO is a next-generation liquid-scintillator reactor neutrino experiment under construction in Southern China. In this scintillator, the two main reactions are the IBD and the 
pES ($\nu +p \rightarrow \nu'+p$) which is possible for all flavours. Although the cross-section of the process is three times smaller than the IBD one, it is the channel that reports the greatest number of events, because of the contribution of all six flavours neutrinos. 
The complete cross-section for this process can be written as \citep{Weinberg:1972,Ahrens:1987}:
\begin{equation}
    \frac{d\sigma}{dE^\prime}(E_\nu,E^\prime)=\frac{G_F^2 m_p}{2\pi E_\nu^2}\left[(C_V\pm C_A)^2 E_\nu^2+(C_V \mp C_A)(E_\nu-E^\prime)^2 -(C_V^2-C_A^2)m_pE^\prime\right]\,,
\end{equation}
where $E^\prime$ and and $E_\nu$ are the proton recoil energy and the incoming neutrino energy. The upper (lower) sign corresponds to neutrinos (antineutrinos). 
$C_V=1/2-2\sin^2\theta_w$ and $C_A=\frac{g_A(0)(1+\eta)}{2}$ are the vector and axial-vector coupling constants respectively; $\theta_w$ is the effective weak mixing angle ($\sin^2 \theta_w = 0.23155$), $g_A(0)\sim 1.26$ is the axial proton form factor \citep{Beringer:2012}, and $\eta$ is the proton strangeness, which is the contribution of the s-quark to $g_A(0)$ ($\eta=0.12\pm0.07$) \citep{Ahrens:1987}. 
For this channel,  ${E_\nu^{min}}=\frac{E^\prime+\sqrt{E^\prime(E^\prime+2m_p)}}{2}$ is the minimum neutrino energy required to reach a distinct $E^\prime$.
The visible energy $E$, \emph{i.e.} the energy measured by the detector, is strongly quenched with respect to $E^\prime$. 
To convert the true proton energy spectra into spectra of visible energy $E$, we applied  a proton quenching factor  that follows a Birk's law $E(E^\prime) =\int_0^{E^\prime} \frac{dE^\prime}{1+ kB(dE^\prime/dx) + C(dE^\prime/dx) ^2}$. The values for the  energy-loss rate of protons in
the material ($dE^\prime/dx$), $kB = 0.0098$ cm MeV$^{-1}$, and C\,=\,0 m$^2$\,MeV$^{-2}$ were estimated by \citep{vonkrosigk:2013}.
For this detector we considered a fiducial volume of 18.25 kt with a number of targets of $N_p=1.318e+33$ and an energy resolution of $\delta=3\%\sqrt{E^\prime}$. We setted the threshold $E_{{th}_d}=0.2$ MeV and the cut-off energy $E_{max}=60$ MeV \citep{Wang:2021}. 

\subsection{DUNE}
DUNE will be made up of four 10-kt liquid argon time projection chambers (LArTPCs).
As described in the Technical Design Report \citep{Abi:2020-TDR}, DUNE will have 70-kt liquid ultra-pure $^{40}$Ar mass in total, of which 40 kt will be fiducial mass (10-kt fiducial mass per module). For each module, we have assumed an amount of $N_{ar} = 1.56\times10^{32}$ target argon nuclei, and  $N_e=2.7\times10^{33}$ target electrons \citep{Abi:2020}. We have also considered a detector efficiency and an energy resolution extracted from Ref. \citep{Abi:2021}. 
The dominant reaction in this detector is $\nu_e+^{40}\rm{Ar} \rightarrow e^- + ^{40}\rm{K}^*$, followed by $\bar{\nu}_e+^{40}\rm{Ar} \rightarrow e^+ + ^{40}\rm{Cl}^*$. For this case, we use the MARLEY \citep{Gardiner:2021}  cross-sections implemented in SNOwGLoBES \citep{Scholberg:2021}. 
The threshold $E_{{th}_d}$ and the cut-off energy $E_{max}$ were setted at 5 and 100 MeV respectivley.

\section{Neutrino signal results}\label{sec:result_nu_signal}

We analyze the temporal distribution of events by considering the primary reactions in each of the mentioned detectors across all models in our sample. Additionally, we determine the expected total number of events for each channel, denoted as $\langle{N}\rangle$. To compute the signals, we employ the sntools code Monte Carlo event generator \citep{Abe:2021}, which we have modified to incorporate the JUNO, DUNE, and HK detectors. The calculation incorporates factors such as detector efficiency, smearing effects, and channel-specific thresholds, as described in Section \ref{sec:nu_signal}. 

\begin{figure}[h]
  \centering
        \includegraphics[width=1.\textwidth]{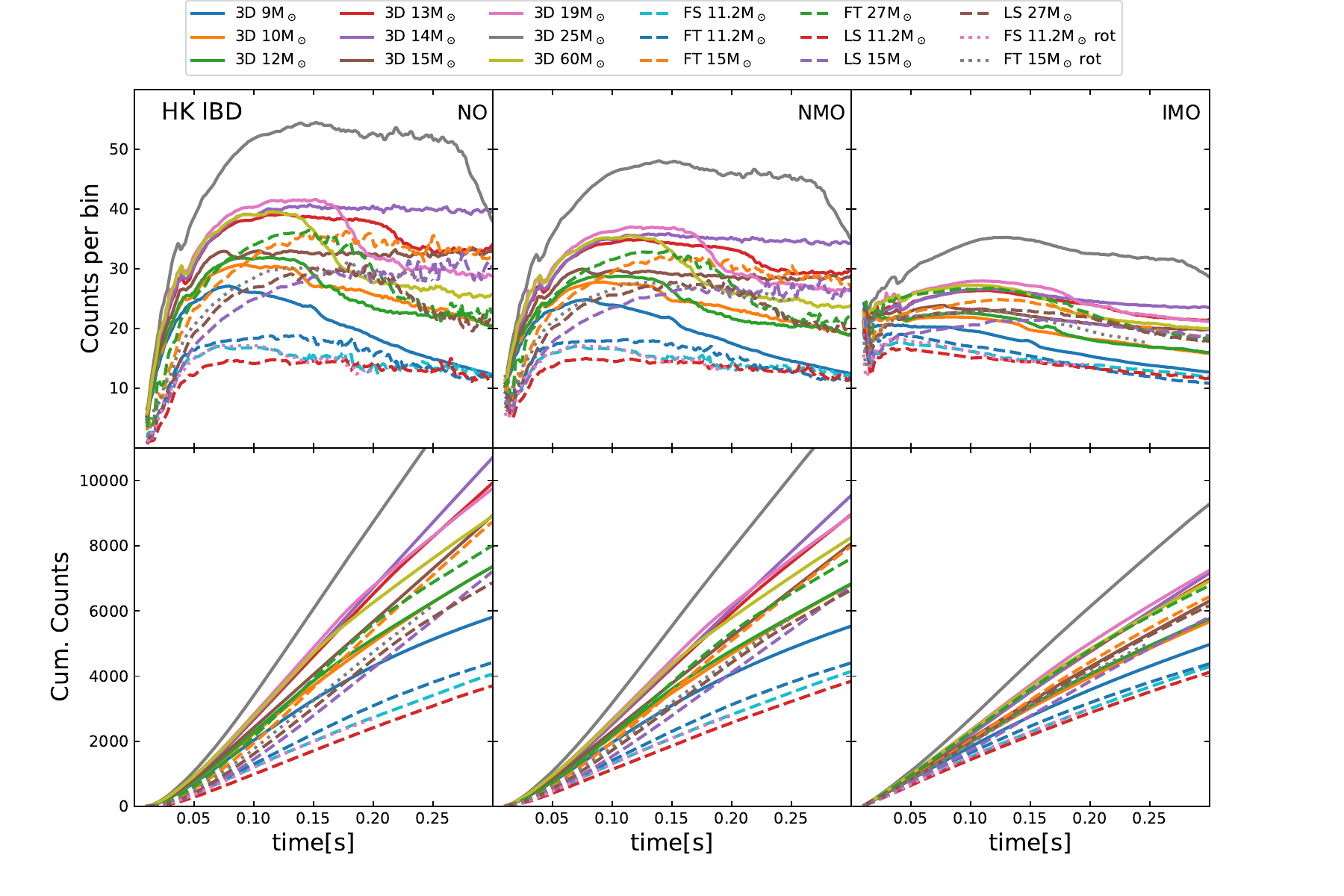} \caption{ First row: expected events per time bin for all the SN models considered for the IBD channel at HK detector. Second row: Cumulative counts for the same models. The first column corresponds to the case without considering neutrino oscillations, while the second and third correspond to NMO and IMO respectively. In all cases, we have assumed a SN distance of 10kpc.
        }\label{time_counts_HK}
    \end{figure}

\begin{figure}[h]
  \centering
        \includegraphics[width=1.\textwidth]{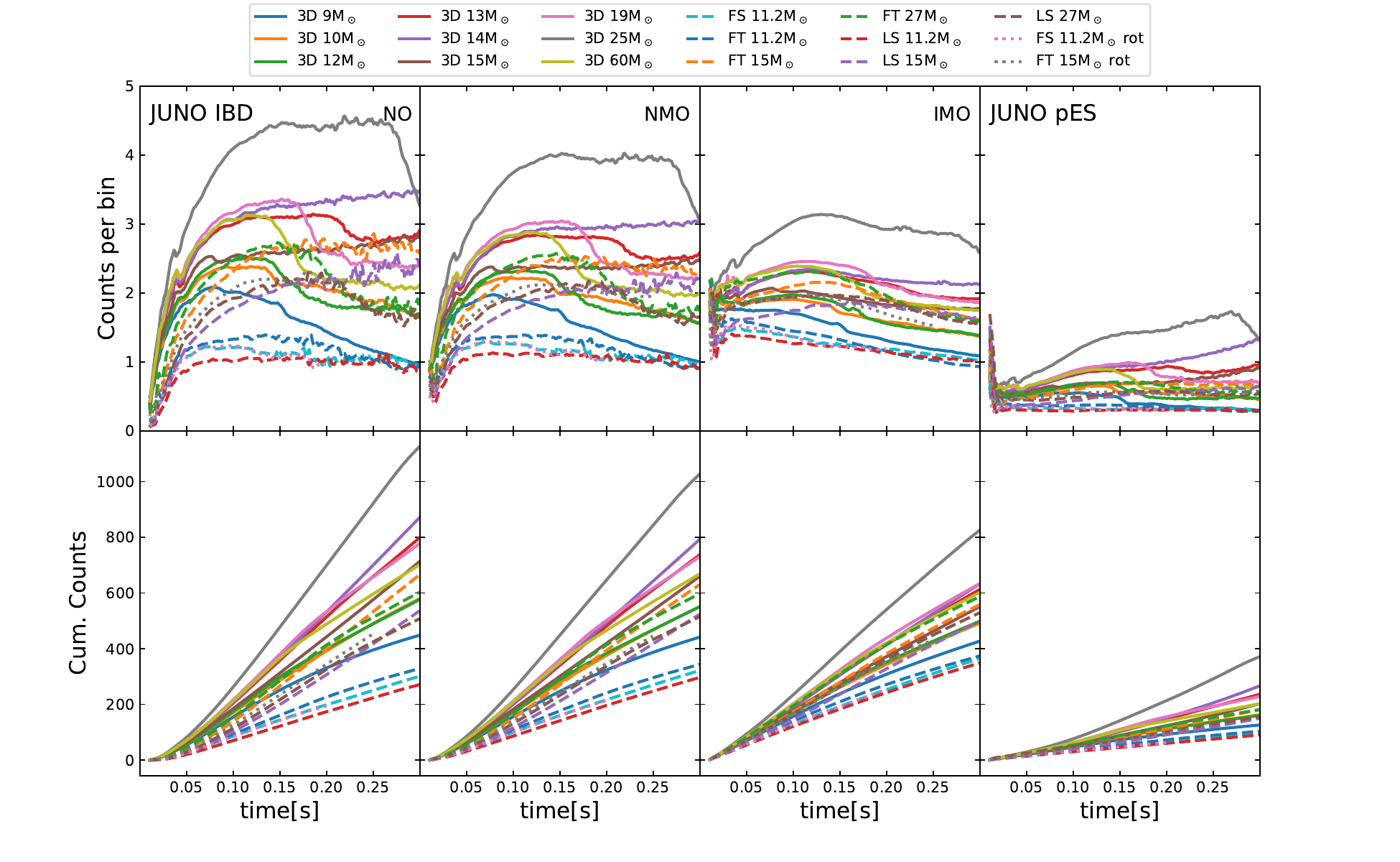} \caption{ Same as Figure \ref{time_counts_HK} but for the IBD and pES channels at JUNE detector.}\label{time_counts_JUNO}
    \end{figure}

\begin{figure}[h]
  \centering
        \includegraphics[width=1.\textwidth]{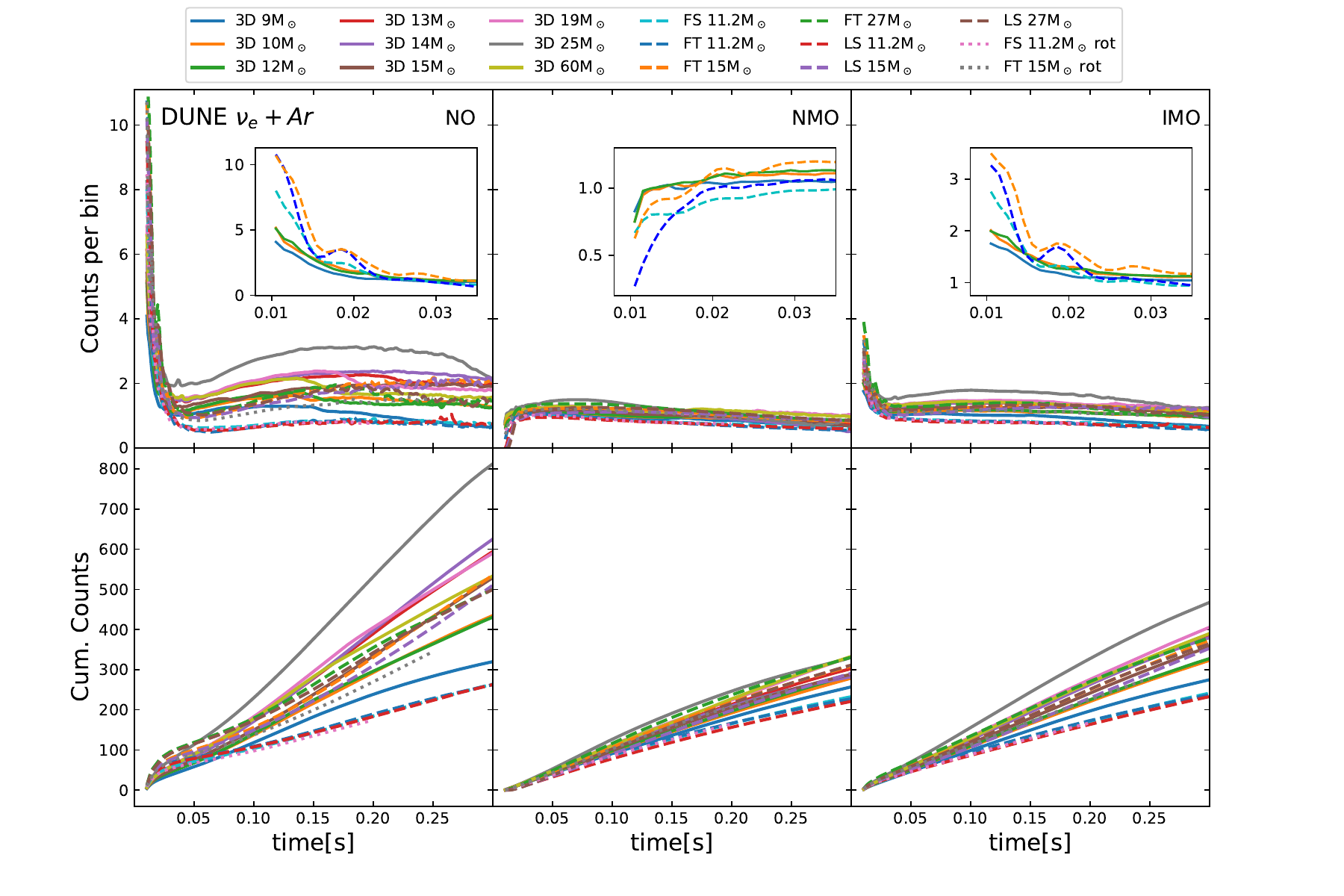} \caption{Same as Figure \ref{time_counts_HK} but for the $\nu_e + Ar$ channel at DUNE detector. In addition, in a zoomed-in view we show the neutronization peak for the first three models from each set: 3D 9 M$_\odot$, 3D 10 M$_\odot$, and 3D 12 M$_\odot$, as well as FS 11.2 M$_\odot$, FT 11.2 M$_\odot$, and FT 15 M$_\odot$. } \label{time_counts_DUNE}
    \end{figure}

In Figures \ref{time_counts_HK}, \ref{time_counts_JUNO}, and \ref{time_counts_DUNE}, we present the expected counts per bin (cpb) (first row) and the cumulative counts (second row) for the HK, JUNO and DUNE detectors respectively. The first columns correspond to the case without oscillations (NO), while the second and third columns correspond to NMO and IMO, respectively. For the JUNO detector, an additional fourth column is included corresponding to the neutral pES channel. In all the cases, we considered a SN distance of 10kpc and a final integration time of 0.3 seconds since it is a time that all the models considered reach. 

For the IBD channel in both HK and JUNO detectors, the 3D models typically yield a higher event count compared to the events generated by the 2D models. This is related to the behavior of the spectral parameters shown in the second row of Fig. \ref{spectral_param_vs_t}. The $\bar{\nu}_e$ average energy and luminosities  are consistently greater in 3D models compared to their 2D counterparts. Moreover, the $\beta$ factor exhibits a systematic reduction in 3D models in contrast to the 2D models, leading to increased event rates within the IBD channel of HK and JUNO.  An exception is the 3D model with 9 M$_\odot$, which has reduced luminosities and energies due to an early lack of an accretion component in the neutrino emission (see \cite{Nagakura:2021} for details). Also, the three 2D models with masses of 11.2\,M$_\odot$ have the fewest number of events due to the lowest luminosities and mean energies as mentioned above. 
This would indicate that the neutrinosphere associated with that neutrino flavor of the 2D models is located at a greater radius and lower temperature than in the 3D models, an effect that reduces the energy of the $\bar{\nu}_e$. 
In the case with NMO, the electron antineutrinos have a high probability of survival ($\sim 67 \%$), thus the trend observed for NO in the IBD channels is maintained, although more modest.  For the IMO case, the difference between the models for the IBD channel in HK and JUNO is smaller, since the counts in this case reflect mostly the properties of the $\bar{\nu}_x$. As expected, the pES channel does not present alterations due to the inclusion of oscillations.

\begin{figure}[h]
  \centering
        \includegraphics[width=0.6\textwidth]{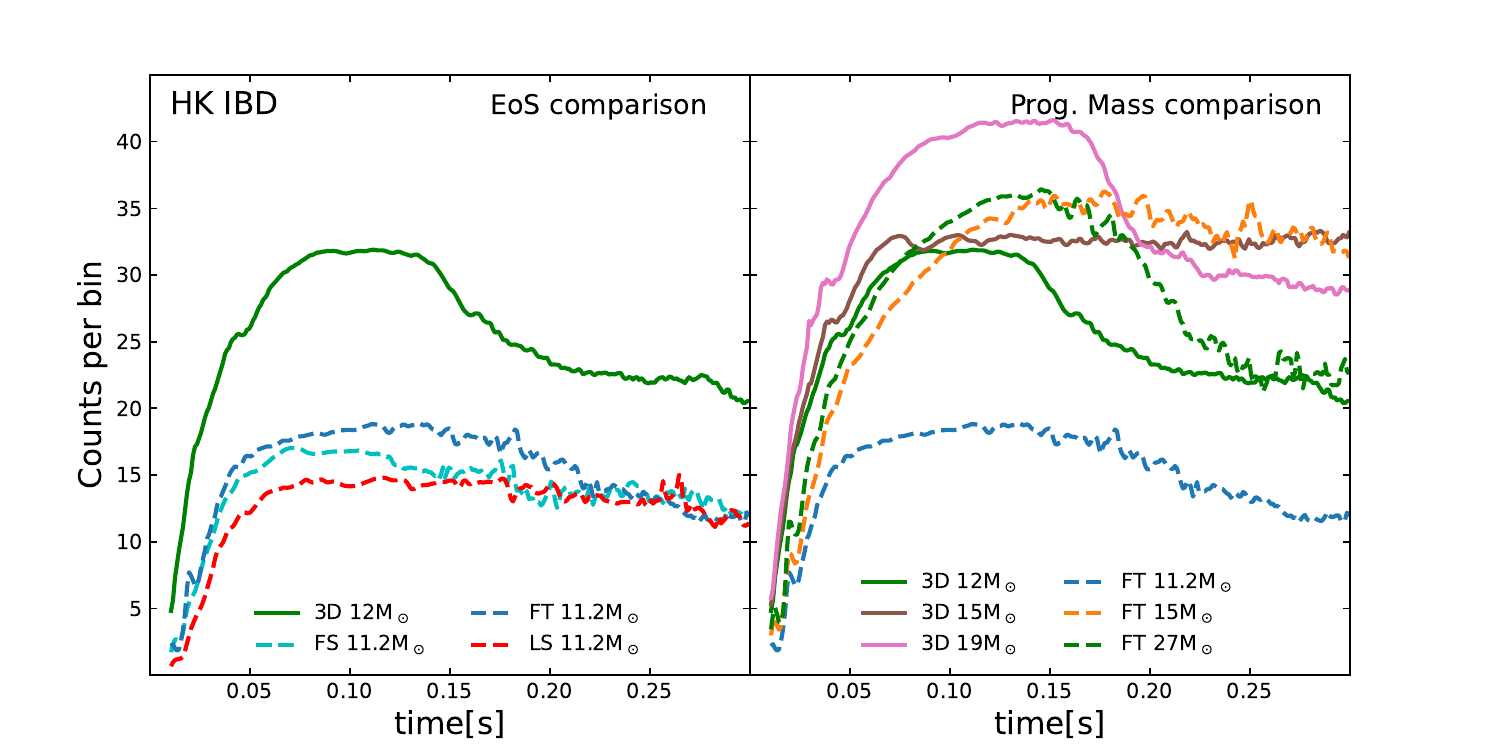} \caption{ Expected events per time bin for some of the SN models considered for the IBD channel at HK detector for the no oscillation case. Left panel: comparison of models with the same progenitor mass, but different EoS. Right panel: Comparison of models with the same EoS but different progenitor mass.}\label{time_counts_HK_2}
    \end{figure}

In Figure \ref{time_counts_HK_2}, we present a reduced number of models to better visualize the differences resulting from changes in EoS and progenitor mass for the IBD channel. In the left panel, we compare models with same progenitor masses but different EoS. It is observed that $\rm{cpb(LS)} < \rm{cpb(FS)} < \rm{cpb(FT)}$, reflecting the hierarchy observed in the $\bar{\nu_e}$ luminosities, as described in Section \ref{spectral_info}. Additionally, we include a 3D model that uses a SFHo EoS, which produces more counts than the others.
In the right panel, we observe changes when varying the progenitor mass while keeping the same EoS. Generally, a greater progenitor mass correlates with higher counts. However, as previously observed with luminosities, this relationship is not strictly monotonic and varies based on the compactness of each model. For the JUNO detector, the trends remain the same since it involves the same interaction channel.

Regarding the signals at DUNE for the NO case, we also found that the 3D models exhibit a higher event count compared to the 2D models. Again, this is attributed to the behavior of the spectral parameters of the electron neutrinos, wherein the 3D models demonstrate higher energies and luminosities compared to the 2D models. When considering NMO, the peak of the neutronization burst is significantly suppressed. This phenomenon arises due to the nearly negligible probability of survival for $\nu_e$, resulting in the DUNE signal being predominantly influenced by the characteristics of $\nu_x$. Given that $L_{\nu_x}$ is lower than $L_{\nu_e}$, the event rate decreases compared to scenarios without flavor conversions. In the IMO case, the neutronization burst persists, albeit partially suppressed, as the probability of survival for $\nu_e$ stands at approximately 30 percent.
The differences in this early signal between various oscillation schemes are more pronounced in the 2D models than in the 3D ones. When comparing the early signal between NMO and IMO, the differences in the 3D models are modest (around a 2.5 factor), whereas in the 2D models, the results differ by a factor greater than four.
In the inset plots of Figure \ref{time_counts_DUNE}, we display a zoomed-in view of the neutronization peak for three models from each set: 3D 9 M$_\odot$, 3D 10 M$_\odot$, and 3D 12 M$_\odot$, as well as FS 11.2 M$_\odot$, FT 11.2 M$_\odot$, and FT 15 M$_\odot$.

\begin{figure}[h]
  \centering
        \includegraphics[width=.6\textwidth]{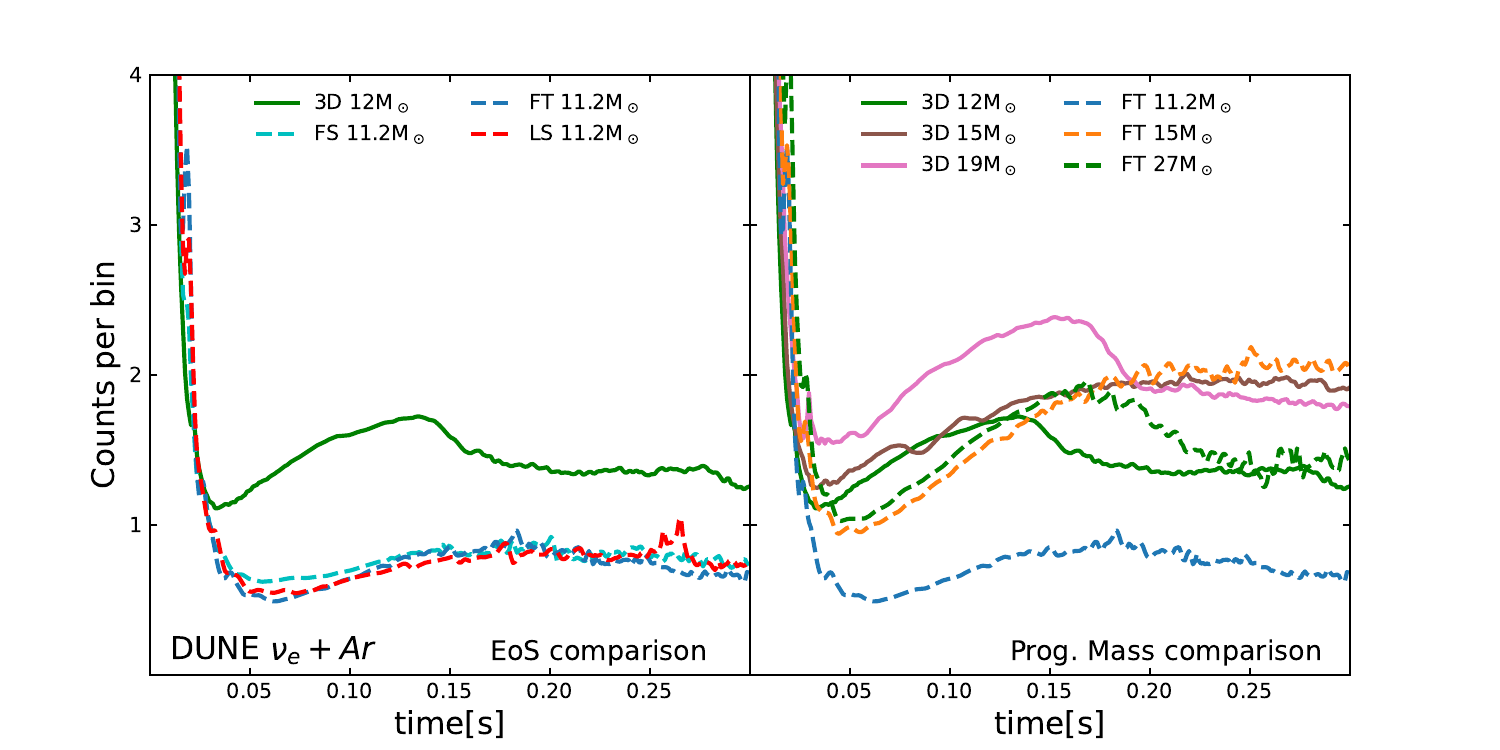} \caption{ Expected events per time bin for some of the SN models considered for the $\nu_e + Ar$ channel at DUNE detector for the no oscillation case. Left panel: comparison of models with the same progenitor mass, but different EoS. Right panel: Comparison of models with the same EoS but different progenitor mass..}\label{time_counts_DUNE_2}
    \end{figure}

In \ref{time_counts_DUNE_2}, we present a reduced number of models in order to better visualize the effects of considering different EoS or progenitor masses for DUNE signals. Once again, the behavior is related to the effects that different EoS or characteristics of the initial model have on compactness. Greater compactness leads to a smaller neutrinosphere radius and, consequently, higher neutrinosphere energies. For the illustrated examples $\rm{cpb(FT)} < \rm{cpb(LS)} < \rm{cpb(FS)}$, mirroring the behavior observed in the $\nu_e$ luminosities.

Finally, we calculated the total expected events in each channel for all the models under study. We observed that the model generating the least number of counts in the four channels studied is the 2D FS 11.2 M$_\odot$, producing approximately $\sim 4000$ counts in HK. Conversely, the model yielding the highest counts in the four channels is the 3D 25 M$_\odot$, generating approximately $\sim 13700$ counts in HK. The values for all the models and channels analyzed are presented in Table \ref{table:events_real} in appendix \ref{appB}. These mean values will be important for later defining our probability distributions and likelihoods in Section \ref{sec:model_discr}.


\subsection{TONE and cumulative events correlation}

 In the previous sections we saw that the time evolution of both event rate and cumulative event number strongly depends on the progenitor characteristics. It is interesting to look for theoretical correlations between observed quantities (e.g., the number of events at each detector) and the total neutrino energy (TONE) based on the selected  models, which are expected to be less sensitive to the progenitor as examined by \cite{Nagakura:2021} for the 3D models.  TONE is defined as energy- and flavor-integrated time-cumulative neutrino radiation in the unit of energy up to a given post-bounce time. The post-bounce time dependence of TONE for all our models is given in Fig. \ref{timevstone} in Appendix \ref{appB}.

 In Figures \ref{tone_counts_real_HK},  \ref{tone_counts_real_JUNO}, and \ref{tone_counts_real_DUNE}, we present the cumulative counts observed in each detector as a function of TONE for NO, NMO, and IMO. The incorporation of a more extensive set of models in our study has allowed us to confirm that there exists a universal correlation between the cumulative number of events in different reaction channels and TONE. Thanks to this correlation, we can estimate TONE from observed neutrino data at each detector by assuming a neutrino oscillation model. It is worth mentioning that we extended the correlation for the first time to the Boltzmann-radiation-hydrodynamics models. A significant finding here is that the robust correlation across different progenitors exists not only for the 3D models but also for the 2D models, though the relations for the 2D models themselves differ from the 3D counterpart. Notably, even among the 2D models, which employ different equations of state and rotation effects, the correlation remains remarkably consistent.

The mass-hierarchy dependence of the correlation implies that the correlation is primarily driven by the heavy-lepton neutrinos $\nu_x$ originating from the source. The correlation found in Figures \ref{tone_counts_real_HK},  \ref{tone_counts_real_JUNO} and \ref{tone_counts_real_DUNE}  gets tighter when oscillations are taken into account. Because the oscillations convert the $\nu_x$ at the source into $\nu_{\rm e}$ and $\bar{\nu}_{\rm e}$ at the detector, this means that the count--TONE correlation is stronger for $\nu_x$ than $\nu_{\rm e}$ and $\bar{\nu}_{\rm e}$. This is also understood from the fact that $\nu_x$ has the largest contribution to the total neutrino luminosity, the integrand of the TONE, as $L=L_{\nu_{\rm e}} + L_{\bar{\nu}_{\rm e}} + 4 L_{\nu_x}$ where the factor 4 comes from four species of neutrinos represented by $\nu_x$. The degree of correlation is influenced by the survival probability of the neutrinos, further emphasizing the significance of including oscillations and heavy-lepton neutrinos in the analysis of the observed data.

In Figure \ref{tone_counts_real_HK}, we present the correlation for the HK IBD channel. We observe that the 3D models exhibit higher counts for the same TONE value in comparison to the 2D models, resulting in a gap between the two groups of simulations. This can be attributed to the behavior of the spectral parameters of the electron antineutrinos, where the 3D models exhibit higher luminosity and lower pinching parameters compared to the 2D models. When considering the effects of oscillations, the heavier lepton neutrinos gain significance. Due to their display of similar spectral parameters across both sets of models, this leads to a reduction in the observed gap Figure \ref{tone_counts_real_JUNO} exhibits a similar pattern, as it also concerns the IBD channel, but for JUNO detector. The last column displays the correlation for the neutral pES channel, which exhibits a smaller gap. This is primarily because this channel is inherently sensitive to heavy neutrinos, and the oscillations do not play a significant role.

We have observed a narrower correlation among all models in the context of DUNE, as depicted in Figure \ref{tone_counts_real_DUNE}, with a smaller gap between the 2D and 3D models. While it is challenging to precisely determine the reason for this effect, it can be attributed to the significance of the neutronization burst in the overall signal. We observed that, for the NO scenario, the neutronization peak was more pronounced in the 2D models than in the 3D models, contributing to achieving a balance in the total accumulated counts between the two sets of models. Also, although the exact reasons remain uncertain and are beyond the scope of this paper, it is possible that the high-energy tail of the energy spectra formed by the PNS surface structure \citep{Keil:2003} and possibly the shock acceleration \citep{NagakuraHotokezaka:2021} may play some roles because the $\nu_e +$ Ar channel is sensitive to the high-energy neutrinos as evidenced in Figure \ref{xs}. In addition, for this channel the $\nu_x$  at the source turn into $\nu_e$ at the Earth almost completely (${P}_e\sim$ 0 or 0.3 for NMO and IMO respectively), indicating that DUNE would provide the most sensitive data with which to measure the TONE. The correlation serves a dual purpose by enabling us to determine the most accurate model for describing the observed supernova and extracting essential parameters like the proto-neutron star  mass and radius \citep{Nagakura:2022}.  

In addition to confirming the correlation reported by Ref. \citep{Nagakura:2021} but for the 2D models, it is important to emphasize that within the examined data set, the correlation remains consistent regardless of the progenitor mass, the choice of equation of state, and the inclusion or exclusion of rotation. This independence further strengthens the reliability and applicability of the correlation in the broader context of supernova modeling.

 \begin{figure}[h]
  \centering
        \includegraphics[width=0.8\textwidth]{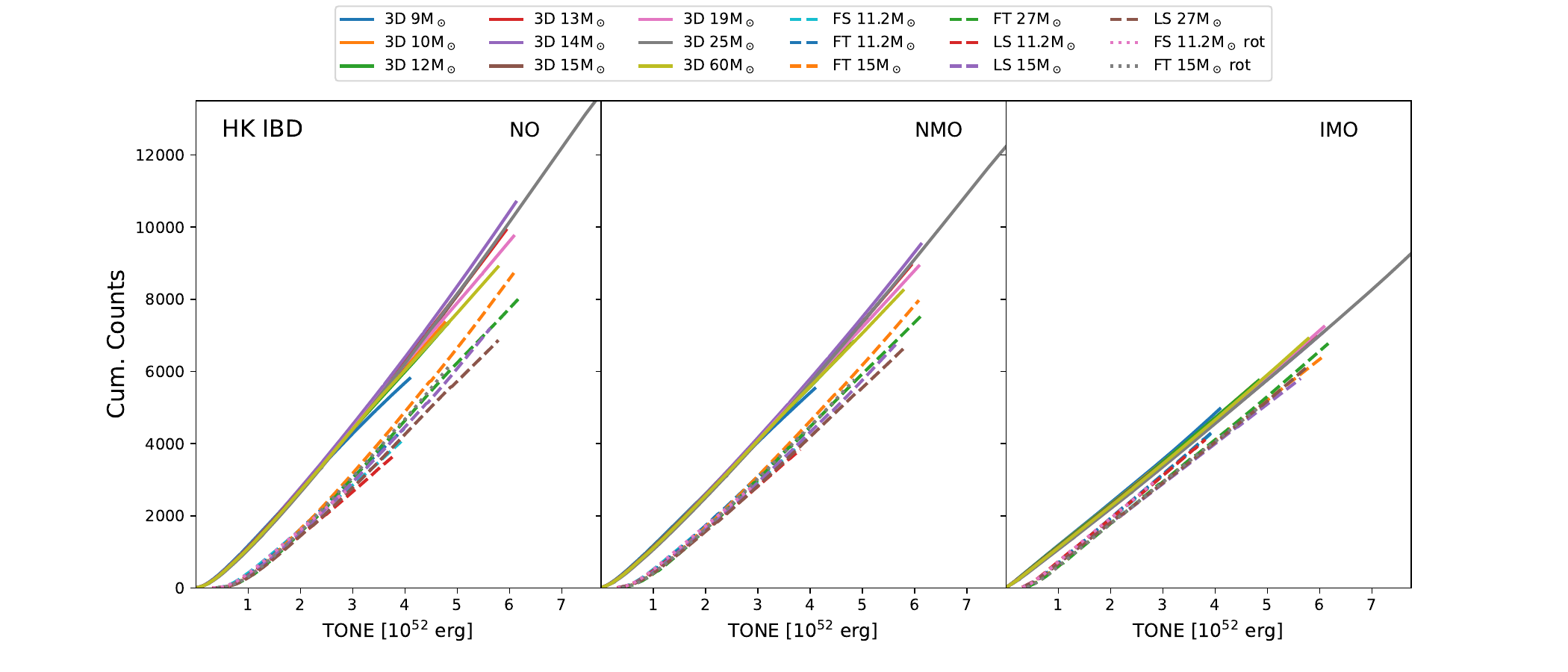} \caption{ Cumulative counts for all the studied models as a function of the total neutrino energy (TONE) for the HK detector. The first column corresponds to the case without considering neutrino oscillations (NO), while the second and third correspond to NMO and IMO respectively.}\label{tone_counts_real_HK}
    \end{figure}

\begin{figure}[h]
  \centering
        \includegraphics[width=1.0\textwidth]{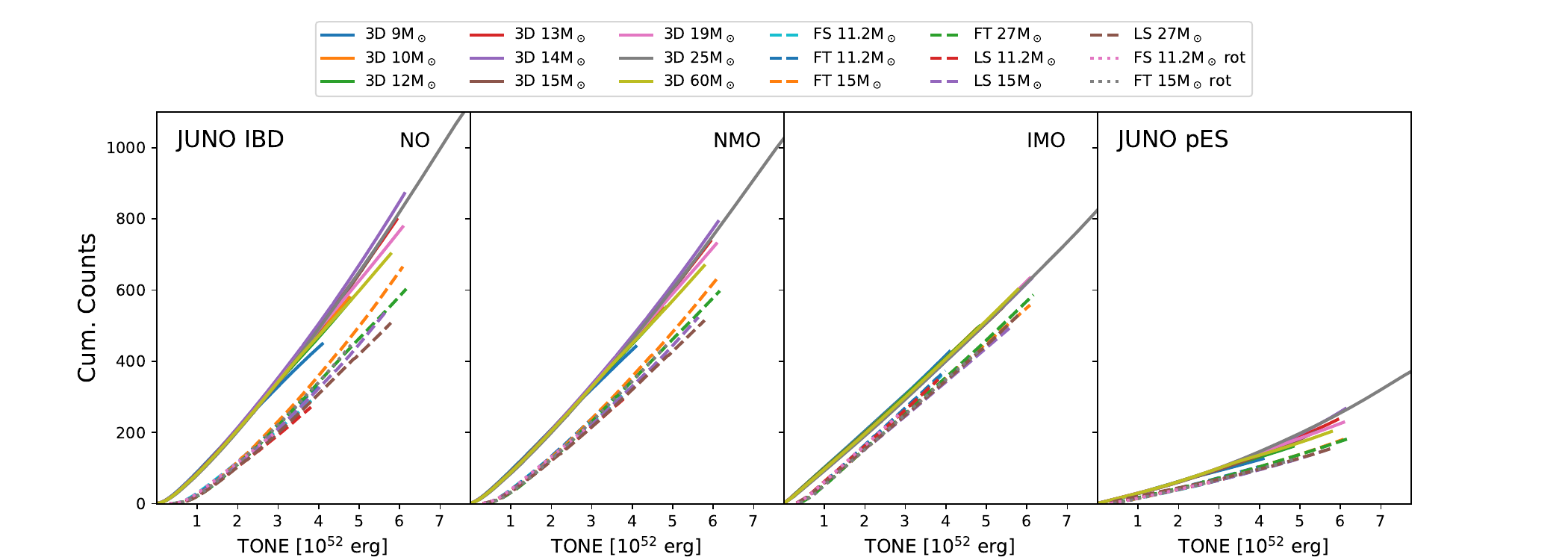} \caption{ Same as Figure \ref{tone_counts_real_HK} ut for JUNO detector. The first three columns correspond to the IBD channel, while the fourth to the pES channel. }\label{tone_counts_real_JUNO}
    \end{figure}
    
\begin{figure}[h]
  \centering
        \includegraphics[width=0.8\textwidth]{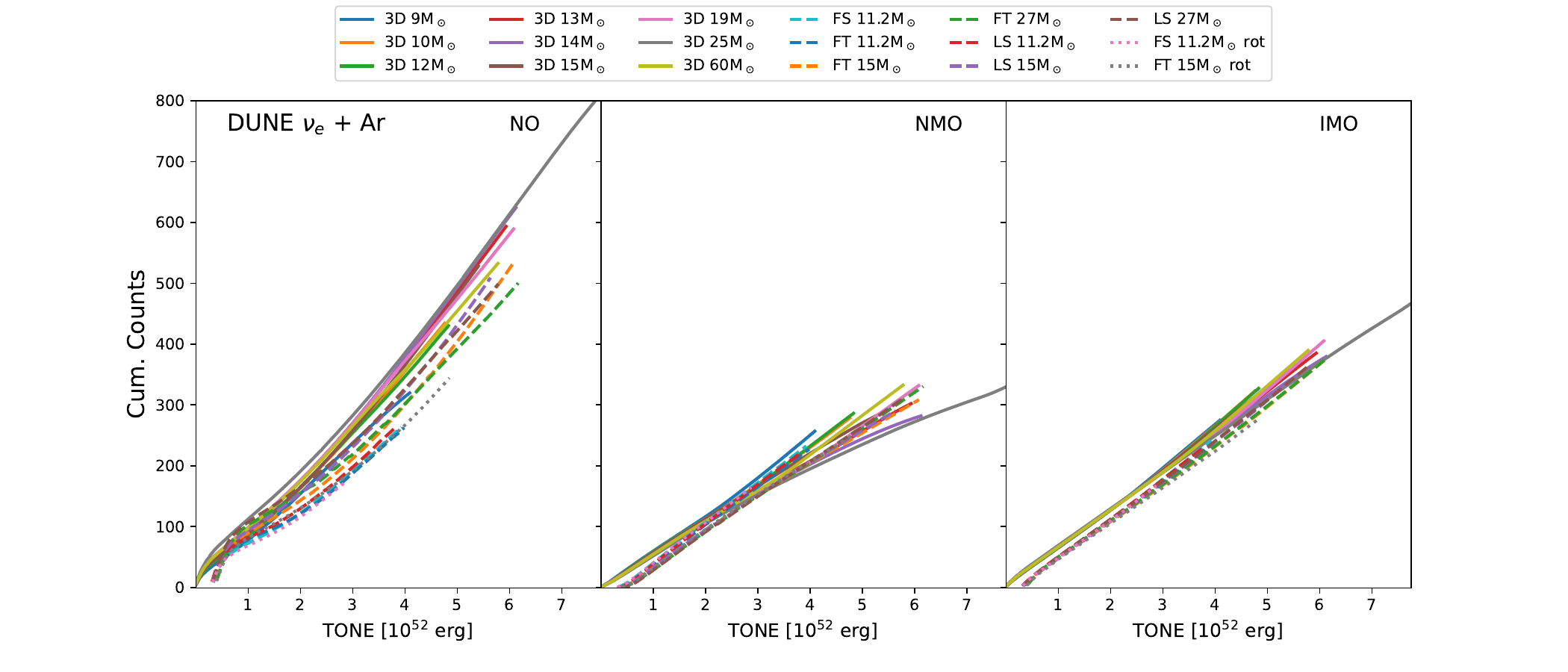} \caption{  Same as Figure \ref{tone_counts_real_HK} but for DUNE detector.}\label{tone_counts_real_DUNE}
    \end{figure}

In Appendix \ref{appB}, we provide approximate formulae for calculating the correlation in realistic scenarios encompassing both neutrino oscillation models. These formulas could be valuable tools for estimating the TONE derived from real observations of cumulative counts, in particular, we found quadratic expressions in the cases with NMO and linear ones in the cases with IMO. Since the correlation shows dispersion depending on whether 3D or 2D models are used, we give a fit for each set of simulations.  
In the case of the DUNE detector,  the simulations conducted using different models across various dimensions have exhibited improved agreement. Consequently, a unified approach is employed, where a single fit is performed using all the available data.  This channel is particularly advantageous as it exhibits a higher level of independence from the models employed, making it a reliable choice for estimating TONE.

We have refined the existing fits for the 3D simulations given by \cite{Nagakura:2021} by incorporating new detection channels, and considerations such as fiducial volumes, efficiencies, and smearing effects specific to our detectors\footnote{The fittings found differ from those given in \citep{Nagakura:2021}, since there, the counts have been calculated using the SNOwGLoBES software, incorporating smearing matrices. In our case, we have modeled the smearing effects using the Gaussian functions described in Sec. \ref{sec:nu_signal}. Also, the detector configurations (fiducial volumes, efficiencies, etc) are not exactly the same in both works. We have not used SNOwGLoBES, because for the statistical analysis, we not only need a mean event rate calculator, but also an associated Monte Carlo event generator.}. Additionally, we have expanded the range of models used in the fitting process to encompass a greater variety.   

When we reconstruct TONE from cumulative number of events, we need to specify a neutrino oscillation model, for which we assumed a simple neutrino oscillation model (adiabatic MSW). On the other hand, neutrino-neutrino self-interaction gives rise to another class of oscillations, known as collective neutrino oscillations. During early times such as the neutronization burst and accretion phase, it is expected that the matter potential dominates over the neutrino--neutrino potential, strongly suppressing self-induced effects, in particular, the associated with the so-called \emph {slow} transformations, that can induce spectral swaps and splits \citep{Chakraborty:2011,Sarikas:2011}. In such cases, the flavor evolution of neutrinos is determined solely by the influence of matter, as assumed along this work.  Nevertheless, the previously mentioned suppression has faced recent challenges \citep{Bhattacharyya:2022,Richers:2021,Wu:2021,Dasgupta:2015}. Several investigations propose that temporal instabilities in the dense neutrino gas can enable self-induced effects, even when a dominant matter density is present \citep{Dasgupta:2015}. This is called the \emph{fast} flavor conversion (FFC) \citep{Chakraborty:2016, Sawyer:2016}. The FFC takes place quite universally in the supernova core \citep{Nagakura:2021b, Harada:2022, Morinaga:2022}, and it modifies the neutrino spectra and supernova dynamics through neutrino heating \citep{Ehring:2023, Nagakura:2023}. 
Despite attempts to characterize these effects, our current understanding of them remains far from settled due to its dynamic range from the oscillation scale $\mathcal{O}({\rm m})$ to the system size $\mathcal{O}(10^3\,{\rm km})$. Furthermore, the \emph{collisional} instability has been gathering attention very recently \citep{Johns:2021}, making the situation more puzzling. This highlights the crucial significance of exploring neutral channels like pES and their correlation with TONE. These channels possess sensitivity to all neutrino flavors and remain unaffected by the chosen oscillation scheme.

\section{Supernova Model Discrimination}\label{sec:model_discr}
In this section, following the ideas developed by \cite{Olsen:2022}, \cite{Abe:2021}, and  \cite{Migenda:2020}, we perform a statistical analysis aimed at evaluating the viability of distinguishing between different SN neutrino emission models within our data set.  The models featuring rotation will be excluded from this analysis due to the limited diversity in progenitor masses and EoS, which makes meaningful comparisons unfeasible. We have taken into account the three aforementioned detectors and their primary reaction channels. The pES channel in the JUNO detector have not exhibited distinguishability due to its low count rate. Therefore, it is not included in the following analysis. For each model, we consider the standard MSW effect, incorporating either the normal or inverted neutrino mass hierarchy, alongside the reference scenario of no neutrino oscillations. 

For a specific emission model $M_j$, the probability distribution for an event to be observed at time t with energy E is
\begin{equation}
    p(E,t|M_j)=\frac{1}{\braket{N}}\frac{d^2N}{dtdE}
\end{equation}
We start by calculating the expected number of events, $\braket{N}$, for each model, as shown in Table \ref{table:events_real}. Subsequently, we generate a set $D=\{E_i, t_i | i = 1, 2, · · · , N\}$ by randomly sampling N events from the distribution, where $E_i$ denotes the energy and $t_i$ represents the emission time of the ith event.

Following the extended maximum likelihood function definition  from \cite{Barlow:1990},  we consider the likelihood of a simulated signal to be 

\begin{equation}
\mathcal{L}(D|M_j)=\frac{e^{-\braket{N}} \braket{N}^N}{N!}\prod_{i=1}^Np(E_i,t_i|M_j) 
\end{equation}
In this case, this is also the bayesian evidence, so the Bayes factor that compares the distribution of neutrino energy and time for models $\alpha$ and $\beta$ results:

\begin{equation}
\mathcal{B}_{\alpha\beta}=\frac{\mathcal{L}(D|M_\alpha)}{\mathcal{L}(D|M_\beta)}
\end{equation}
The last can be used to determine whether $M_\alpha$ is favored over
$M_\beta$ given the data D. For convenience, we use the natural logarithm of the Bayes factor.  The criteria for interpreting $\ln\mathcal{B}_{{\alpha\beta}}$ is give in Refs. \citep{Loredo:2002, Kass:1995}. The larger $\ln\mathcal{B}_{{\alpha\beta}}$ is, the more strongly $M_\alpha$ is favored over $M_\beta$.

\begin{table}[ht!]
\begin{center}
\caption{Bayes factor interpretation as suggested in \citep{Loredo:2002,Kass:1995}. }
\vspace{0.5cm}
{\renewcommand{\arraystretch}{1}
\begin{tabular}{ccc}
\hline
 $\ln\mathcal{B}_{\alpha\beta}$&  Interpretation\\ \hline
0 to 1 &  Not worth more than a bare mention\\
1 to 3 &  Positive  evidence favoring M$_\alpha$\\
3 to 5 & Strong evidence favoring M$_\alpha$\\
$>$ 5 &  Very Strong evidence favoring M$_\alpha$\\
\hline
\end{tabular}}\label{table:ln_bayes_factor_interpretation}
\end{center}
\end{table}

By executing this procedure a total of $10^4$ times and leveraging our Monte Carlo simulated signals, we are able to compute both the mean $\langle \ln\mathcal{B}_{{\alpha\beta}} \rangle$ and the standard deviation $\sigma (\ln\mathcal{B}_{{\alpha\beta}})$ across all pairs of our model instances. For this simple likelihood, the calculation of the mean and standard deviation can also be derived analytically using the expressions presented by \cite{Olsen:2022}. We utilize these expressions to verify the consistency of the results. Additionally, the Monte Carlo simulation has the potential to incorporate other free parameters in future analyses. To discern a substantial preference for $M_\alpha$ over $M_\beta$, we establish a threshold of $\ln\mathcal{B}_{\alpha\beta} > 5$ (as detailed in Table \ref{table:ln_bayes_factor_interpretation}).

For a more stringent measure of model differentiation, we take into account the approximately normal distribution of $\ln\mathcal{B}_{{\alpha\beta}}$ (see appendix \ref{appA}). We declare models $M_\alpha$ and $M_\beta$ to be distinguishable at a 95\% confidence level (CL) if the condition $\langle \ln\mathcal{B}_{{\alpha\beta}} \rangle - 1.96\sigma (\ln\mathcal{B}_{{\alpha\beta}})  > 5$ is satisfied. In the upcoming subsections, we will present the averages $\overline{\ln\mathcal{B}_{\alpha,\beta}} = \frac{1}{2}(\ln\mathcal{B}_{\alpha\beta} + \ln\mathcal{B}_{\beta\alpha})$ as a means of both symmetrizing and summarizing our findings. For a detailed explanation of the calculation process behind these averages, please refer to Appendix \ref{appA}.

\subsection{EoS discrimination} 

This analysis aims to discern the impact of varying equations of state on the model outcomes. By calculating Bayes factors, we can assess the strength of evidence for one model over another. However, since the 3D models under examination share the same EoS (SFHo), they do not contribute relevant insights to this particular comparison. We compute the Bayes factors  between the 2D models that share identical progenitor, but differ in their implemented EoS.  The results are presented in Table \ref{table:ln_bayes_factor_EoS},  the bold entries  indicates that the true model $M_\alpha$ can be distinguished from the alternative model $M_\beta$ at the 95\% CL.  In the tables we show the symmetrized results, that is, averaging between the results found when calculating $\ln\mathcal{B}_{{\alpha\beta}} $ and  $\ln\mathcal{B}_{{\beta\alpha}} $.
In the appendix \ref{appA}, we show in detail the symmetrization procedure for one of the entries.

\begin{table}[ht!]
\begin{center}
\caption{The mean Bayes factors along with standard deviations $\langle \overline{\ln\mathcal{B}_{{\alpha,\beta}}} \rangle \pm \sigma (\overline{\ln\mathcal{B}_{{\alpha,\beta}}})$ for a SN distance of 10kpc regarding the EoS discrimination. The two models can be distinguished at the  95\% CL
for $\langle \overline{\ln\mathcal{B}_{{\alpha,\beta}}} \rangle - 1.96\sigma (\overline{\ln\mathcal{B}_{{\alpha,\beta}}})  > 5$ for all the entries depicted in bold.} 
\vspace{0.5cm}
\tiny{\begin{tabular}{ccc|cc|cc}
\toprule
\textbf{ HK  IBD }&  & & \textbf{DUNE $\nu_e$ + Ar}& &  \textbf{ JUNO  IBD } \\
\toprule
$M_\alpha / M_\beta$ & FT 11.2M$_\odot$  & LS 11.2M$_\odot$ &  FT 11.2M$_\odot$  & LS 11.2M$_\odot$ & FT 11.2M$_\odot$  & LS 11.2M$_\odot$\\ \hline
FS 11.2M$_\odot$&    && &  & & \\
NO & \textbf{39.34 $\pm$ 8.62} &  \textbf{56.65 $\pm$ 10.29}  & {3.51 $\pm$ 2.69} &{2.81 $\pm$ 1.89} & {3.71 $\pm$ 2.72} &  {4.52 $\pm$ 3.57}
\\
NMO  &   \textbf{27.38 $\pm$ 6.97} & \textbf{27.42 $\pm$ 8.14}  &  {0.55 $\pm$ 1.18 }& {0.82 $\pm$ 0.86} & {2.05 $\pm$ 1.46} &  {2.28 $\pm$ 3.06}\\
IMO   &  \textbf{ 22.17 $\pm$ 6.81} & {18.48 $\pm$ 7.44}  &  {0.65 $\pm$ 1.49} & {0.94 $\pm$ 0.93} & {2.02 $\pm$ 1.48} & 1.18 $\pm$ 0.38
 \\
FT 11.2M$_\odot$ &   &&& && \\
NO   &  { } & \textbf{141.75 $\pm$ 16.08} &  & {2.34 $\pm$ 1.76} & & {11.14 $\pm$ 7.85}
 \\
NMO &   {} &  \textbf{76.38 $\pm$ 12.53} & &   {0.64 $\pm$ 0.57} & & {6.40 $\pm$ 6.67}
\\
IMO   &  {}  &  \textbf{29.58 $\pm$ 7.78} & & {0.84 $\pm$ 0.59} & & {2.72 $\pm$ 5.30}
\\

\hline
$M_\alpha / M_\beta$ &   LS 15M$_\odot$ &  & LS 15M$_\odot$ &  & LS 15M$_\odot$\\ \hline
FT 15M$_\odot$&    && &&\\
NO  &  \textbf{278.94 $\pm$ 22.49} &  & {2.89 $\pm$ 2.63} & & {23.97 $\pm$ 10.48}\\
NMO   &  \textbf{171.71 $\pm$ 17.93} &   & {1.01 $\pm$ 2.15} & & {15.26 $\pm$ 9.64}\\
IMO   &  \textbf{73.47 $\pm$ 12.40} &   & {1.71 $\pm$ 2.16} & & {6.12 $\pm$ 6.82}\\
\hline
$M_\alpha / M_\beta$  & LS 27M$_\odot$ &   & LS 27M$_\odot$ &   & LS 27M$_\odot$\\ \hline
FT 27M$_\odot$&    && &&\\
NO &   \textbf{216.62 $\pm$ 20.14}    &  & {2.37 $\pm$ 2.25} & & {18.39 $\pm$ 9.95} \\
NMO &    \textbf{139.11 $\pm$ 17.05} &   & {1.24 $\pm$ 2.07}  & & {11.26 $\pm$ 9.28}\\
IMO &    \textbf{71.25 $\pm$ 12.41} &   & {1.28 $\pm$ 2.04} & & {6.26 $\pm$ 6.78}\\
\end{tabular}}\label{table:ln_bayes_factor_EoS}
\end{center}
\end{table}

From the  values presented in Table \ref{table:ln_bayes_factor_EoS}, it is evident that at a distance of $10$ kpc, all simulations with varying EoS exhibit distinguishable characteristics at the 95\% CL compared to the other models considered, when observed at the HK detector. The most noticeable differentiation within the IBD channel, is observed when comparing FT and LS models. This observation aligns with shown in Figure \ref{time_counts_HK_2}. The only case that does not yield distinguishability occurs in the case of FS and LS EoS, specifically when analyzing IMO and a progenitor mass of $11.2$ M$_\odot$. This outcome can be attributed to the striking similarity in the temporal distribution of electron antineutrinos between these two models.

Moreover, the level of distinguishability concerning the equation of state is noticeably influenced by the adopted mass ordering. It is noteworthy that the scenario involving the IMO presents the least sensitivity among the cases considered.

Upon investigating the signal within the context of the DUNE experiment, it becomes evident that the ability to differentiate between different EoS scenarios diminishes. This aligns with what is shown in Figure \ref{time_counts_DUNE_2}, where it can be seen that the effects of different EoS on the DUNE signal are small. This can be attributed to two primary factors: reduced count statistics and the striking resemblance in spectral distributions and luminosities of the electron neutrino flavor between the various 2D models, particularly when they share the same progenitor mass. As observed previously, DUNE stands out as the detector that consistently exhibits the strongest correlation in count numbers across all models. This observation aligns with the understanding that the challenge of distinguishing between the different models becomes more pronounced in this context. Regarding the JUNO detector, while its sensitivity surpasses that of DUNE, it falls short of providing sufficient capability to effectively distinguish between different models. The resemblance in behavior based on the mass ordering mirrors what is observed in HK, as both detectors depend on the same interaction channel, but with lower counts and statistics.

\subsection{Progenitor mass discrimination} 
Here we explore the distinguishability based on the mass of the progenitor. To achieve this, we compare models sharing identical EoS and calculated using the same code.

The results for the 2D models for the three detectors are presented in Table \ref{table:ln_bayes_factor_prog_mass1}. In the context of HK, all the models are distinguishable from each other; however, in the case of DUNE, the power of distinguishability decreases. These findings align with those depicted in Figures \ref{time_counts_HK_2} and \ref{time_counts_DUNE_2}, specifically in the right panels where the effects on the signal due to variations in the progenitor's mass are illustrated.

The results for the 3D models are displayed in Tables \ref{table:ln_bayes_factor_prog_mass2} and \ref{table:ln_bayes_factor_prog_mass3} and \ref{table:ln_bayes_factor_prog_mass4} for HK, DUNE and JUNO respectively. In the case of HK, all the models are distinguishable from each other, except for the models with 10 and 12 M$_\odot$. Once again, within this set of models, the distinguishability power in DUNE is insufficient to achieve a 95\% confidence level in most cases. Across all instances, it can be observed that the inclusion of oscillations complicates the distinguishability between models. Regarding HK, the scenario with NMO exhibits a higher discriminatory capability among models. Conversely, in the case of DUNE, this distinction is more prominent for IMO. This is related to what is depicted in the top panel of Figure \ref{time_counts_DUNE}, where it can be observed that the event distribution in DUNE is more similar across all models for the NMO scenario. In JUNO, the behavior closely resembles that of HK but with reduced event counts, thus diminishing the discrimination capability. The 3D 25 M$\odot$ model exhibits the highest distinguishability among the scenarios, possibly influenced by the fact that all initial models for this set of simulations were computed by \citep{Sukhbold:2016}, except for the 25 M$_\odot$ progenitor, which was derived from \citep{Sukhbold:2018}. Consequently, when comparing scenarios involving this model, the distinguishability increases not only due to variations in progenitor mass but also because of differences in the initial model itself.

\begin{table}[ht!]
\begin{center}
\caption{The mean Bayes factors along with standard deviations $\overline{\langle \ln\mathcal{B}_{{\alpha,\beta}}} \rangle \pm \sigma (\overline{\ln\mathcal{B}_{{\alpha,\beta}}})$  for a SN distance of 10kpc, regarding the progenitor-mass discrimination  for the 2D models of our sample. The two models can be distinguished at the  95\% CL
for $\langle \overline{\ln\mathcal{B}_{{\alpha,\beta}}} \rangle - 1.96\sigma (\overline{\ln\mathcal{B}_{{\alpha,\beta}}})  > 5$ for all the entries, except for those not in bold. }
\vspace{0.5cm}
{\renewcommand{\arraystretch}{1}
\tiny{\begin{tabular}{ccc|cc|cc}
\toprule
&\textbf{ HK  IBD}& & \textbf{DUNE $\nu_e$ + Ar}& & \textbf{JUNO IBD} &  \\
\toprule
$M_\alpha / M_\beta$ &  FT 15M$_\odot$  & FT 27M$_\odot$&FT 15M$_\odot$  & FT 27M$_\odot$ & FT 15M$_\odot$  & FT 27M$_\odot$ \\ \hline
FT 11.2M$_\odot$& &   & & \\
NO &    \textbf{1572.51 $\pm$ 59.41} &  \textbf{1099.66 $\pm$ 47.85}  & \textbf{114.30 $\pm$ 16.75}  & \textbf{83.90 $\pm$ 13.52} & \textbf{126.98 $\pm$ 17.94} & \textbf{83.41 $\pm$ 15.15}
\\
NMO &      \textbf{1159.26 $\pm$ 48.98} &  \textbf{882.97 $\pm$ 42.88} & {13.84 $\pm$ 6.31} & \textbf{20.24 $\pm$ 7.16} & \textbf{94.51 $\pm$ 15.84} & \textbf{70.99 $\pm$ 13.63}
\\
IMO &   \textbf{467.46 $\pm$ 31.33} & \textbf{559.28 $\pm$ 34.08} & \textbf{34.74 $\pm$ 8.89} &\textbf{34.30 $\pm$ 8.75} & \textbf{43.18 $\pm$ 11.11} & \textbf{50.89 $\pm$ 11.82} \\
FT 15M$_\odot$ & &   &&\\
NO &  &  \textbf{178.36 $\pm$ 19.33}   & & {13.79 $\pm$ 5.46} & & {15.28 $\pm$ 7.85
}\\
NMO &  &   \textbf{ 97.01 $\pm$ 14.01} & & {2.89 $\pm$ 1.40} &  &{8.07 $\pm$ 7.50}\\
IMO &  &   \textbf{22.03 $\pm$ 6.53} & & {2.06 $\pm$ 2.00} & & {4.13 $\pm$ 3.08}\\
\hline
$M_\alpha / M_\beta$ &  LS 15M$_\odot$ & LS 27M$_\odot$&  LS 15M$_\odot$ & LS 27M$_\odot$ &  LS 15M$_\odot$ & LS 27M$_\odot$  \\ \hline
LS 11.2M$_\odot$&  & &&  \\
NO &  \textbf{1227.67 $\pm$ 52.72} &  \textbf{994.21 $\pm$ 46.24} & \textbf{95.42 $\pm$ 14.77} &  \textbf{87.39 $\pm$ 14.56} & \textbf{94.90 $\pm$ 15.96} & \textbf{75.47 $\pm$ 14.02}
\\
NMO &   \textbf{885.34 $\pm$ 44.49} &   \textbf{785.81 $\pm$ 40.99} &{11.04 $\pm$ 5.59} & {16.73 $\pm$ 6.47} & \textbf{69.47 $\pm$ 14.02}
& \textbf{61.21 $\pm$ 13.04}

 \\
IMO &   \textbf{342.00 $\pm$ 26.36}  &   \textbf{443.39 $\pm$ 29.72} & \textbf{28.49 $\pm$ 7.95} & \textbf{30.66 $\pm$ 8.33} & \textbf{31.29 $\pm$ 10.23} & \textbf{39.74 $\pm$ 11.10}

\\
LS 15M$_\odot$ &   & &&  \\
NO &  &   \textbf{107.34 $\pm$ 14.90 } & & {11.42 $\pm$ 4.86} & & {8.63 $\pm$ 6.66
} \\
NMO &   &   \textbf{64.12 $\pm$ 11.33} & & {1.49 $\pm$ 1.04} && {6.18 $\pm$ 5.07}\\
IMO &   &   \textbf{25.66 $\pm$ 7.07} & & {2.07 $\pm$ 1.99} & & {4.04 $\pm$ 2.21}\\
\hline
\end{tabular}}\label{table:ln_bayes_factor_prog_mass1}}
\end{center}
\end{table}

\begin{rotatetable}
\begin{center}
\tablecomments{The mean Bayes factors along with standard deviations $\langle \overline{\ln\mathcal{B}_{{\alpha,\beta}}} \rangle \pm \sigma (\overline{\ln\mathcal{B}_{{\alpha,\beta}}})$  for a SN distance of 10kpc, regarding the progenitor-mass discrimination  for the 3D models of our sample. The two models can be distinguished at the  95\% CL
for $\langle \overline{\ln\mathcal{B}_{{\alpha,\beta}}} \rangle - 1.96\sigma (\overline{\ln\mathcal{B}_{{\alpha,\beta}}})  > 5$ for all the entries, except for those not in bold. }
\vspace{0.5cm}
{\renewcommand{\arraystretch}{1}
\tiny{\begin{tabular}{ccccccccc}
\toprule
\textbf{ HK  IBD}&& & \\
\toprule
$M_\alpha / M_\beta$ & 3D 10M$_\odot$  & 3D 12M$_\odot$ & 3D 13M$_\odot$ & 3D 14M$_\odot$  & 3D 15M$_\odot$ &  3D 19M$_\odot$ & 3D 25M$_\odot$  & 3D 60M$_\odot$\\ \hline
3D 9M$_\odot$ & &   & &   & &  &    \\
NO  & \textbf{280.43 $\pm$ 24.27}  & \textbf{246.35 $\pm$ 22.38} &  \textbf{1390.90 $\pm$ 56.04} & \textbf{1943.07 $\pm$ 70.00}  & \textbf{981.31 $\pm$ 47.97} & \textbf{1166.78 $\pm$ 50.17} & \textbf{2292.04 $\pm$ 60.94} & \textbf{727.75 $\pm$ 38.72}  \\
NMO  & \textbf{189.91 $\pm$ 19.98} & \textbf{175.46 $\pm$ 18.94} & \textbf{1024.18 $\pm$ 48.34} & \textbf{1419.89 $\pm$ 59.28} &\textbf{680.76 $\pm$ 39.34} & \textbf{918.07 $\pm$ 44.20} & \textbf{2073.38 $\pm$ 86.26} & \textbf{590.94 $\pm$ 35.12}\\
IMO  & \textbf{67.06 $\pm$ 11.76} &  \textbf{73.49 $\pm$ 12.14} &  \textbf{461.30 $\pm$ 32.03} & \textbf{618.34 $\pm$ 38.68} & \textbf{243.70 $\pm$ 23.03} & \textbf{504.23 $\pm$ 32.69} & \textbf{1638.47 $\pm$ 64.10} & \textbf{357.51 $\pm$ 26.66} \\
3D 10M$_\odot$ & & & &   \\
NO & & \textbf{17.53 $\pm$ 6.01} & \textbf{466.80 $\pm$ 30.69} & \textbf{835.34 $\pm$ 42.47} & \textbf{241.81 $\pm$ 22.74} &  \textbf{370.38 $\pm$ 27.24}  & \textbf{2195.03 $\pm$ 70.97} & \textbf{171.15 $\pm$ 18.36}\\
NMO &  & {10.99 $\pm$ 4.81} &  \textbf{361.37 $\pm$ 27.84} & \textbf{629.84 $\pm$ 37.79} & \textbf{170.00 $\pm$ 19.19}&  \textbf{315.86 $\pm$ 25.51} & \textbf{1812.26 $\pm$ 63.69} & \textbf{154.29 $\pm$ 17.63} \\
IMO &   &  {5.51 $\pm$ 3.39} & \textbf{188.28 $\pm$ 20.02} &   \textbf{309.79 $\pm$ 26.73} & \textbf{65.32 $\pm$ 11.94}& \textbf{218.43 $\pm$ 21.10} & \textbf{1106.25 $\pm$ 50.71} & \textbf{133.42 $\pm$ 16.06}\\
3D 12M$_\odot$ & & & &   \\
NO  &  &  &\textbf{528.85 $\pm$ 33.10} & \textbf{919.08 $\pm$ 45.32} & \textbf{301.92 $\pm$ 25.58} &  \textbf{386.09 $\pm$ 27.89}  & \textbf{2290.79 $\pm$ 74.03} &  \textbf{152.24 $\pm$ 17.34} \\
NMO &  &  &   \textbf{392.90 $\pm$ 28.97}& \textbf{674.41 $\pm$ 39.15} & \textbf{203.12 $\pm$ 20.89} & \textbf{318.24 $\pm$ 25.34} &  \textbf{1853.75 $\pm$ 64.91} & \textbf{138.18 $\pm$ 16.53}\\
IMO &   &   & \textbf{188.44 $\pm$ 20.34} & \textbf{315.18 $\pm$ 27.34} &  \textbf{71.81 $\pm$ 12.63} & \textbf{202.61 $\pm$ 20.45} & \textbf{1092.95 $\pm$ 51.16} & \textbf{113.20 $\pm$ 14.88}\\
3D 13M$_\odot$ & & & &   \\ 
NO &  &  &  & \textbf{78.06 $\pm$ 12.74} & \textbf{82.11 $\pm$ 12.22} &\textbf{66.56 $\pm$ 11.92}  & \textbf{702.07 $\pm$ 37.46} &  \textbf{212.06 $\pm$ 21.07}   \\
NMO & & &   & \textbf{56.54 $\pm$ 10.90} & \textbf{67.34 $\pm$ 11.35} &  \textbf{42.81 $\pm$ 9.44}  & \textbf{598.39 $\pm$ 34.80}  & \textbf{131.52 $\pm$ 16.86}\\
IMO & &  &   &  \textbf{30.55 $\pm$ 7.90} & \textbf{46.49 $\pm$ 9.64} & \textbf{26.83 $\pm$ 7.39} & \textbf{407.86 $\pm$ 29.13}  & \textbf{52.51 $\pm$ 10.58}\\
3D 14M$_\odot$ & & & &   \\
NO & &  & &  & \textbf{215.41 $\pm$ 20.13} & \textbf{247.49 $\pm$ 23.71} &  \textbf{412.43 $\pm$ 28.16}  &  \textbf{492.10 $\pm$ 33.21}\\
NMO &  &  &   &   & \textbf{171.55 $\pm$ 18.57} & \textbf{166.25 $\pm$ 19.01} & \textbf{370.80 $\pm$ 27.14} & \textbf{323.84 $\pm$ 26.92} \\
IMO &   &   &  &  & \textbf{102.13 $\pm$ 14.51} & \textbf{91.74 $\pm$ 14.15} & \textbf{297.41 $\pm$ 24.47} & \textbf{141.62 $\pm$ 17.93}\\
3D 15M$_\odot$ & & & &   \\
NO  &  &  & &  &  & \textbf{148.39 $\pm$ 17.29}  &  \textbf{1155.14 $\pm$ 48.64} & \textbf{173.27 $\pm$ 18.82} \\
NMO &  &  &   &  & & \textbf{114.82 $\pm$ 14.92} & \textbf{991.88 $\pm$ 44.82} & \textbf{107.86 $\pm$ 14.82}\\
IMO &   &   &  &   & & \textbf{94.19 $\pm$ 13.65} & \textbf{693.22 $\pm$ 38.39} & \textbf{64.98 $\pm$ 11.23} \\
3D 19M$_\odot$ & & & &   \\
NO & &   &  &   &  & & \textbf{923.26 $\pm$ 45.41} & \textbf{85.46 $\pm$ 13.09}\\
NMO & & &   & & & & \textbf{693.22 $\pm$ 38.39} & \textbf{57.81 $\pm$ 10.85}\\
IMO & &  &   &  & & & \textbf{414.05 $\pm$ 30.46}  & \textbf{24.89 $\pm$ 7.15}\\
3D 25M$_\odot$ & & & &   \\
NO & &   &  &   &  & & &  \textbf{1437.57 $\pm$ 58.10}\\
NMO & & &   & & & & & \textbf{1103.76 $\pm$ 49.57}\\
IMO & &  &   &  & & & & \textbf{581.59 $\pm$ 36.80}  \\
\hline
\end{tabular}}\label{table:ln_bayes_factor_prog_mass2}}
\end{center}
\end{rotatetable}

\begin{rotatetable}
\begin{center}
\tablecomments{Same as Table \ref{table:ln_bayes_factor_prog_mass2}, but for the  $\nu_e$ + Ar channel at DUNE detector. }
\vspace{0.5cm}
{\renewcommand{\arraystretch}{1}
\tiny{\begin{tabular}{ccccccccc}
\toprule
\textbf{ DUNE $\nu_e$ + Ar}&& & \\
\toprule
$M_\alpha / M_\beta$ & 3D 10M$_\odot$  & 3D 12M$_\odot$ & 3D 13M$_\odot$ & 3D 14M$_\odot$  & 3D 15M$_\odot$ &  3D 19M$_\odot$ & 3D 25M$_\odot$  & 3D 60M$_\odot$ \\ \hline
3D 9M$_\odot$ & &   & &   & &  &    \\
NO  &  \textbf{28.67 $\pm$ 8.27} & \textbf{23.96 $\pm$ 7.36} & \textbf{104.20 $\pm$ 16.98} & \textbf{126.69 $\pm$ 19.80} & \textbf{75.79 $\pm$ 14.60} & \textbf{94.33 $\pm$ 15.26} &  \textbf{243.06 $\pm$ 27.15} & \textbf{62.02 $\pm$ 11.84}\\
NMO  & {2.27 $\pm$ 1.01} & { 2.48 $\pm$ 1.85 } & {5.20 $\pm$  4.02 } & { 5.22 $\pm$0.12 } & {3.71 $\pm$ 2.05} & {4.46 $\pm$ 2.20} & {9.63 $\pm$ 4.17} & 10.41 $\pm$ 5.53\\ 
IMO  & {5.15 $\pm$ 3.57} & {5.52 $\pm$ 3.54} & \textbf{20.68 $\pm$ 7.91} & {18.45 $\pm$ 8.49} & {12.75 $\pm$ 6.26}  & \textbf{27.01 $\pm$ 8.41} & \textbf{50.71 $\pm$ 13.81} & \textbf{20.98 $\pm$ 7.09}\\
3D 10M$_\odot$ & & & &   \\
NO & &  {1.54 $\pm$ 1.99} & \textbf{27.75 $\pm$ 8.21} & \textbf{40.97 $\pm$ 10.53}  & {13.47 $\pm$ 5.91} & \textbf{25.30 $\pm$ 7.45} & \textbf{122.87 $\pm$ 17.79} & {12.18 $\pm$ 4.94}\\
NMO &  & { 0.60 $\pm$ 0.25 } & {3.13 $\pm$ 1.42}  & { 3.23 $\pm$ 0.55} & {1.65 $\pm$  0.46 } & {5.02 $\pm$ 4.07} & {7.20 $\pm$ 5.45 } & {4.92 $\pm$ 3.43}\\
IMO &   &  {0.78 $\pm$ 0.29} & {5.63 $\pm$ 4.40} & {4.19 $\pm$ 5.01} & {2.75 $\pm$ 1.13 } & {9.53 $\pm$ 4.93} & {24.92 $\pm$ 10.21} & {6.37 $\pm$ 3.72} \\
3D 12M$_\odot$ & & & &   \\
NO  &  &  & \textbf{33.35 $\pm$ 9.17} & \textbf{47.51 $\pm$ 11.54} & {18.27 $\pm$ 6.95} & \textbf{28.07 $\pm$ 7.95} & \textbf{131.66 $\pm$ 18.74} & {11.82 $\pm$ 4.97}\\
NMO &  &  &  {2.92 $\pm$ 0.74 } & {3.03  $\pm$  0.96}  & {1.51 $\pm$ 0.25 } & {3.78 $\pm$ 3.82} & { 6.96 $\pm$ 4.01} & {3.64 $\pm$ 3.14}\\
IMO &   &   & {4.99 $\pm$ 4.41}  &  {5.04 $\pm$ 3.56} & {2.82 $\pm$ 1.38 } & {8.60 $\pm$ 4.84} & {23.30 $\pm$ 10.16} & {5.54 $\pm$ 3.55} \\
3D 13M$_\odot$ & & & &   \\ 
NO &  &  &  &  {2.87 $\pm$ 2.55 } & {5.45 $\pm$ 3.41} & {3.18 $\pm$ 2.92} & \textbf{36.61 $\pm$ 9.52} & {10.81 $\pm$ 5.58}\\
NMO & & &   &  {0.73 $\pm$ 0.67}  & { 1.62 $\pm$ 0.49} & {1.84 $\pm$ 1.21} & {4.17$\pm$ 1.51 } & {1.29 $\pm$ 1.12}\\
IMO & &  &   &  {1.09 $\pm$ 0.89 } & {2.03 $\pm$ 1.20} & {0.66 $\pm$ 0.8}  & {7.15 $\pm$ 5.82} & {1.81 $\pm$ 1.31}\\
3D 14M$_\odot$ & & & &   \\
NO & &  & &  & {9.31 $\pm$ 4.89} & {7.94 $\pm$ 5.22} & \textbf{26.73 $\pm$ 7.84}  & {20.37 $\pm$ 7.98}\\
NMO &  &  &   &   &  { 1.75 $\pm$ 0.64 } & {3.26 $\pm$ 1.42} & {4.06 $\pm$ 3.67} & {7.81 $\pm$ 1.44}\\
IMO &   &   &  &  & { 2.57 $\pm$ 0.87} & {1.38 $\pm$ 1.66} & {8.66 $\pm$ 5.23} & {2.52 $\pm$ 1.48}\\
3D 15M$_\odot$ & & & &   \\
NO  &  &  & &  &  &  {10.07 $\pm$ 4.58} & \textbf{64.36 $\pm$ 12.45} & {10.64 $\pm$ 5.12}\\
NMO &  &  &   &  &   & {3.26 $\pm$ 2.59} & {5.67 $\pm$ 3.06} & {3.45 $\pm$ 2.09}\\
IMO &   &   &  &   & & {3.44 $\pm$ 2.77}  & {14.61 $\pm$ 7.75} & {2.15 $\pm$ 2.14} \\
3D 19M$_\odot$ & & & &   \\
NO & &   &  &   &  & &  \textbf{55.53 $\pm$ 10.79} & {4.13 $\pm$ 3.51}\\
NMO & & &   & & & &  {3.43 $\pm$ 0.62  } & {0.99 $\pm$ 0.13}\\
IMO & &  &   &  & & & { 5.53 $\pm$ 4.42} &  {1.57 $\pm$ 0.45}  \\
3D 25M$_\odot$ & & & &  & \\
NO & &   &  &   &  & & & \textbf{63.21 $\pm$ 14.13} \\
NMO & & &   & & & & &{4.15 $\pm$ 1.29}\\
IMO & &  &   &  & & & &  {7.06 $\pm$ 3.85} \\
\hline
\end{tabular}}\label{table:ln_bayes_factor_prog_mass3}}
\end{center}
\end{rotatetable}

\begin{rotatetable}
\begin{center}
\tablecomments{Same as Table \ref{table:ln_bayes_factor_prog_mass2}, but for the IBD channel at JUNE detector. }
\vspace{0.5cm}
{\renewcommand{\arraystretch}{1}
\tiny{\begin{tabular}{ccccccccc}
\toprule
\textbf{ JUNO IBD}&& & \\
\toprule
$M_\alpha / M_\beta$ & 3D 10M$_\odot$  & 3D 12M$_\odot$ & 3D 13M$_\odot$ & 3D 14M$_\odot$  & 3D 15M$_\odot$ &  3D 19M$_\odot$ & 3D 25M$_\odot$  & 3D 60M$_\odot$\\ \hline
3D 9M$_\odot$ & &   & &   & &  &    \\
NO  & \textbf{25.87 $\pm$ 9.30} & \textbf{21.88 $\pm$ 8.61} & \textbf{128.39 $\pm$ 17.89} &\textbf{183.91 $\pm$ 22.00} & \textbf{91.73 $\pm$ 15.19}& \textbf{103.62 $\pm$ 16.09}
& \textbf{345.09 $\pm$ 30.17} & \textbf{62.77 $\pm$ 13.02} \\
NMO  & {17.68 $\pm$ 7.70} & {15.80 $\pm$ 7.70} & \textbf{95.31 $\pm$ 14.79} & \textbf{135.12 $\pm$ 17.98} & \textbf{63.88 $\pm$ 12.65} & \textbf{82.75 $\pm$ 14.10} & \textbf{275.37 $\pm$ 26.02} & \textbf{51.72 $\pm$ 11.46}\\
IMO  & {6.87 $\pm$ 4.21} & {7.40 $\pm$ 4.98} & \textbf{37.61 $\pm$ 7.03} & \textbf{61.74 $\pm$ 12.71} & \textbf{24.14 $\pm$ 8.44}& \textbf{48.14 $\pm$ 11.24} & \textbf{158.38 $\pm$ 19.57} & \textbf{33.23 $\pm$ 9.99}  \\
3D 10M$_\odot$ & & & &   \\
NO & & {5.51 $\pm$ 2.96} &\textbf{42.21 $\pm$ 11.19} & \textbf{79.36 $\pm$ 14.36}& {22.15 $\pm$ 8.88} & \textbf{31.87 $\pm$ 9.99} & \textbf{202.43 $\pm$ 22.22} & {13.75 $\pm$ 7.78} \\
NMO &  &  {4.93 $\pm$ 2.13} & \textbf{33.43 $\pm$ 9.76} & \textbf{60.19 $\pm$ 12.48} & {16.11 $\pm$ 7.61} & \textbf{27.73 $\pm$ 9.20} & \textbf{166.88 $\pm$ 19.95} & {12.84 $\pm$ 6.85}\\
IMO &   &  {3.73 $\pm$ 1.28}& {18.40 $\pm$ 7.81} & \textbf{30.60 $\pm$ 9.16} & {5.75 $\pm$ 3.89} & {20.29 $\pm$ 8.24}& \textbf{105.48 $\pm$ 15.91} & {11.99 $\pm$ 6.86} \\
3D 12M$_\odot$ & & & &   \\
NO  &  &  & \textbf{49.57 $\pm$ 11.66} & \textbf{89.60 $\pm$ 15.17} & \textbf{28.48 $\pm$ 9.86} & \textbf{33.92 $\pm$ 10.60} & \textbf{214.60 $\pm$ 22.74} & \textbf{12.35 $\pm$ 6.98}
\\
NMO &  &  & \textbf{37.27 $\pm$ 10.02} & \textbf{65.76 $\pm$ 13.07} & \textbf{19.54 $\pm$ 8.29} & \textbf{28.58 $\pm$ 9.70} & \textbf{173.62 $\pm$ 20.31} & {11.70 $\pm$ 6.41}\\
IMO &   &   & {18.71 $\pm$ 7.81} & \textbf{31.87 $\pm$ 9.46} & {6.09 $\pm$ 4.07} & {18.91 $\pm$ 8.10} & \textbf{105.75 $\pm$ 16.01} & {10.14 $\pm$ 6.43}\\
3D 13M$_\odot$ & & & &   \\ 
NO &  &  &  &  {7.23 $\pm$ 4.36} & {6.93 $\pm$ 4.98} & {7.14 $\pm$ 4.77} & \textbf{64.52 $\pm$ 13.71} & {19.72 $\pm$ 7.78}  \\
NMO & & &   &  {7.18 $\pm$ 4.19} & {6.60 $\pm$ 3.94} & {6.09 $\pm$ 3.96} & \textbf{54.51 $\pm$ 13.21} & {9.81 $\pm$ 5.35}\\
IMO & &  &   &  {5.06 $\pm$ 3.48} & {5.71 $\pm$ 3.57} & {4.77 $\pm$ 3.15} & \textbf{38.22 $\pm$ 10.38} & {6.22 $\pm$ 3.64}\\
3D 14M$_\odot$ & & & &   \\
NO & &  & &  & {16.23 $\pm$ 7.98} & \textbf{25.13 $\pm$ 9.18} & \textbf{36.47 $\pm$ 11.62} & \textbf{50.74 $\pm$ 11.67} \\
NMO &  &  &   &   & {13.45 $\pm$ 6.55} & {15.52 $\pm$ 6.44} & \textbf{32.45 $\pm$ 10.92} & \textbf{33.01 $\pm$ 9.72}\\
IMO &   &   &  &  & {7.06 $\pm$ 4.36}  & {5.89 $\pm$ 3.89} & \textbf{27.45 $\pm$ 9.53} & {8.09 $\pm$ 4.07} \\
3D 15M$_\odot$ & & & &   \\
NO  &  &  & &  &  &  {12.66 $\pm$ 8.42} & \textbf{105.66 $\pm$ 16.99} & {16.02 $\pm$ 8.69} \\
NMO &  &  &   &  & & {10.02 $\pm$ 7.18} & \textbf{91.65 $\pm$ 15.69} & {8.00 $\pm$ 5.59}\\
IMO &   &   &  &   & & {8.15 $\pm$ 6.62} & \textbf{65.65 $\pm$ 12.99} & {6.02 $\pm$ 4.64}\\
3D 19M$_\odot$ & & & &   \\
NO & &   &  &   &  & & \textbf{89.22 $\pm$ 15.24} & {7.29 $\pm$ 4.58}\\
NMO & & &   & & & & \textbf{69.29 $\pm$ 13.76} & {6.87 $\pm$ 3.58}\\
IMO & &  &   &  & & & \textbf{39.85 $\pm$ 10.17} & {4.81 $\pm$ 3.36} \\
3D 25M$_\odot$ & & & &   \\
NO & &   &  &   &  & & & \textbf{139.92 $\pm$ 18.44} \\
NMO & & &   & & & & & \textbf{106.78 $\pm$ 16.44}\\
IMO & &  &   &  & & & & \textbf{57.52 $\pm$ 11.80} \\
\hline
\end{tabular}}\label{table:ln_bayes_factor_prog_mass4}}
\end{center}
\end{rotatetable}

\subsection{Mass ordering discrimination} 

Lastly, we turn our attention to investigating distinguishability concerning mass ordering. To accomplish this, we perform a comparative analysis between the same simulation, but considering both normal and inverted mass ordering scenarios. 
In the first, second and third columns of the table \ref{table:ln_bayes_factor_mass_ordering}, we display the results obtained for HK, DUNE and JUNO, respectively. We see that at $d = 10$ kpc,
In both HK, DUNE, and JUNO experiments, the majority of scenarios involving can be differentiated with a CL exceeding 95\%. 

\begin{table}[ht!]
\begin{center}
\caption{The mean Bayes factors along with standard deviations $\langle \overline{\ln\mathcal{B}_{{\alpha,\beta}}} \rangle \pm \sigma (\overline{\ln\mathcal{B}_{{\alpha,\beta}}})$  for a SN distance of 10 $\rm kpc$, regarding the neutrino mass ordering discrimination. The two models can be distinguished at the  95\% CL
for $\langle \overline{\ln\mathcal{B}_{{\alpha,\beta}}} \rangle - 1.96\sigma (\overline{\ln\mathcal{B}_{{\alpha,\beta}}})  > 5$ for all the entries depicted in bold. }
\vspace{0.5cm}
\tiny{\begin{tabular}{cc|c|c}
\toprule
&\textbf{ HK IBD } &   \textbf{DUNE $\nu_e$ + Ar} & \textbf{ JUNO IBD }  \\
\toprule
 $M_\alpha / M_\beta$  &  IMO & IMO & IMO \\ \hline
FS 11.2M$_\odot$ &&  &     \\ 
NMO &  \textbf{421.25 $\pm$ 24.84} &  {13.17 $\pm$ 4.78 } & \textbf{36.64 $\pm$ 9.85}\\
FT 11.2M$_\odot$ &&  & \\ 
NMO &   \textbf{484.19 $\pm$ 29.37} & \textbf{17.17 $\pm$ 5.14} &\textbf{41.80 $\pm$ 10.30} \\
FT 15M$_\odot$ &  &  &\\ 
NMO & \textbf{1156.26 $\pm$ 48.89} & \textbf{30.76 $\pm$ 6.50} &\textbf{74.37 $\pm$ 15.90}
\\
FT 27M$_\odot$ & & & \\ 
NMO  &\textbf{953.31 $\pm$ 42.51}&  \textbf{30.05 $\pm$ 6.44} &\textbf{68.66 $\pm$ 14.65}
\\
LS 11.2M$_\odot$ &  & &   \\ 
NMO & \textbf{473.83 $\pm$ 27.32} & \textbf{18.18 $\pm$ 5.24} &\textbf{41.61 $\pm$ 10.49}
 \\
LS 15M$_\odot$ & & & \\ 
NMO &  \textbf{1021.29 $\pm$44.53}  & \textbf{34.29 $\pm$ 6.67} &\textbf{72.24 $\pm$ 14.32}
\\ 
LS 27M$_\odot$ &  &  & \\
NMO & \textbf{914.57 $\pm$ 40.71} & \textbf{35.79 $\pm$ 6.82} &\textbf{69.24 $\pm$ 14.33} \\
3D 9M$_\odot$  &  &  & \\ 
NMO &\textbf{477.65 $\pm$ 30.86} & {9.57 $\pm$ 3.85} &\textbf{30.11 $\pm$ 10.42}
\\
3D 10M$_\odot$  &  &  &  \\ 
NMO & \textbf{626.42 $\pm$ 37.26} &\textbf{14.59 $\pm$ 4.31} &\textbf{39.04 $\pm$ 11.89}
  \\
3D 12M$_\odot$  &  &  &  \\ 
NMO & \textbf{626.27 $\pm$ 35.97} &  \textbf{14.39 $\pm$ 4.35} &\textbf{37.32 $\pm$ 12.22}
 \\
3D 13M$_\odot$  &  &  &  \\ 
NMO & \textbf{946.44 $\pm$ 45.47}&  \textbf{25.37 $\pm$ 5.15} &\textbf{56.49 $\pm$ 14.59} \\
3D 14M$_\odot$  &  &  &  \\ 
NMO & \textbf{1099.72 $\pm$ 50.60} & \textbf{30.97 $\pm$ 5.39} &\textbf{64.07 $\pm$ 15.45}

\\
3D 15M$_\odot$  &  & &   \\ 
NMO &  \textbf{849.89 $\pm$ 43.69}& \textbf{20.82 $\pm$ 4.86} & \textbf{50.80 $\pm$ 13.97}
\\
3D 19M$_\odot$  & & &  \\ 
NMO & \textbf{868.20 $\pm$ 42.97} & \textbf{20.27 $\pm$ 3.58} &\textbf{52.55 $\pm$ 13.73} \\
3D 25M$_\odot$  &  &  &  \\ 
NMO & \textbf{1384.53 $\pm$ 55.17} & \textbf{47.57 $\pm$ 6.33} &\textbf{80.95 $\pm$ 17.60} \\ 
3D 60M$_\odot$  &  & &  \\
NMO & \textbf{766.91 $\pm$ 39.89} & \textbf{19.09 $\pm$ 4.79} &\textbf{45.63 $\pm$ 13.26}
 \\
\end{tabular}}\label{table:ln_bayes_factor_mass_ordering}
\end{center}
\end{table}

\subsubsection{Neutronization bust at DUNE}

Next, we explore the feasibility of utilizing the signal during the neutronization burst in the DUNE experiment to differentiate between mass orderings in Table \ref{table:ln_bayes_factor_mass_ordering_neut_burst}. Our investigation encompasses signals up to 75 milliseconds. 
We observe that even though we only analyze the signal for a short duration of a few milliseconds, DUNE it can still provide distinguishability against mass ordering in certain scenarios. As mentioned earlier, the 2D models exhibit the most significant disparity in the electron neutrino peak, making them the ones with the highest level of distinguishability.

\begin{table}[ht!]
\begin{center}
\caption{The mean Bayes factors along with standard deviations $\langle \overline{\ln\mathcal{B}_{{\alpha,\beta}}} \rangle \pm \sigma (\ln\mathcal{B}_{{\alpha\beta}})$  for a SN distance of 10kpc, regarding the neutrino mass ordering discrimination during the neutronization burst. The two models can be distinguished at the  95\% CL
for $\langle \overline{\ln\mathcal{B}_{{\alpha,\beta}}} \rangle - 1.96\sigma (\overline{\ln\mathcal{B}_{{\alpha,\beta}}})  > 5$ for all the entries depicted in bold.}
\vspace{0.5cm}
\tiny{\begin{tabular}{cc}
\toprule
 \textbf{DUNE $\nu_e$ + Ar}&    \\
\toprule
 $M_\alpha / M_\beta$  &  IMO   \\ \hline
FS 11.2M$_\odot$ &    \\ 
NMO &  {8.49 $\pm$ 2.97}\\
FT 11.2M$_\odot$ &   \\ 
NMO &   \textbf{12.55 $\pm$ 3.08}\\
FT 15M$_\odot$ &   \\ 
NMO &  \textbf{13.80 $\pm$ 3.45}\\
FT 27M$_\odot$ &  \\ 
NMO  & \textbf{17.80 $\pm$ 3.60} \\
LS 11.2M$_\odot$ &    \\ 
NMO & \textbf{13.06 $\pm$ 2.77} \\
LS 15M$_\odot$ &  \\ 
NMO &  \textbf{17.08 $\pm$ 2.86} \\ 
LS 27M$_\odot$ &    \\
NMO & \textbf{21.70 $\pm$ 2.98} \\
3D 9M$_\odot$  &   \\ 
NMO & {6.64 $\pm$ 2.32} \\
3D 10M$_\odot$  &      \\ 
NMO & {8.14 $\pm$ 2.42}  \\
3D 12M$_\odot$  &      \\ 
NMO & {8.20 $\pm$ 2.45} \\
3D 13M$_\odot$  &     \\ 
NMO & 10.58 $\pm$ 2.64\\
3D 14M$_\odot$  &     \\ 
NMO &  {9.88 $\pm$ 2.67}\\
3D 15M$_\odot$  &    \\ 
NMO &  {9.13 $\pm$ 2.56}  \\
3D 19M$_\odot$  &    \\ 
NMO  &\textbf{11.68 $\pm$ 2.76}
\\
3D 25M$_\odot$  &     \\ 
NMO & \textbf{15.24 $\pm$ 3.07} \\ 
3D 60M$_\odot$  &     \\
NMO &  \textbf{11.40 $\pm$ 2.74} \\
\end{tabular}}\label{table:ln_bayes_factor_mass_ordering_neut_burst}
\end{center}
\end{table}
\section{Discussion and conclusions}\label{sec:conclusions}

For the very first time, we have conducted a comprehensive analysis of the anticipated neutrino signal derived from Boltzmann-radiation-hydrodynamics models. This study encompasses a systematic exploration across a range of terrestrial detectors.
Moreover, we delve into a comparative investigation by contrasting our predictions with those previously formulated for the Fornax 3D models. Our examination of 18 models concurrently enables us to uncover disparities in aspects such as temporal event distributions, cumulative count patterns, and spectral profiles. Remarkably, a pivotal finding emerges as we identify a significant correlation between the TONE and the cumulative count statistics. This compelling relation, consistently validated across all our models, is detailed through equations applicable to different channels. We have identified that the connection we observed is primarily influenced by the massive lepton neutrinos that originate from the source. 
The strength of this correlation is shaped by the surviving probabilities, which highlights the crucial aspect of accounting for neutrino oscillations in our analysis. This results underscores not only the significance of massive lepton neutrinos but also the impact of oscillations on the data we have gathered. Additionally, we have found that the correlation between all models is notably narrower when observed through the DUNE detector.
Despite inherent uncertainties within our theoretical frameworks, the robustness of these correlations is poised to yield valuable insights in the interpretation of actual observations.

Finally, we aimed to assess the potential of planned neutrino detectors in discerning between different SN models. By employing Bayesian techniques,  we can effectively rank various SN neutrino emission models in the event of an actual SN occurrence.  Based on the findings presented in this work, we have strong grounds to believe that the forthcoming Galactic SN's neutrino signal will provide us with the means to differentiate among a diverse array of models. This critical capability promises to significantly advance our understanding of these cosmic events and contribute to the broader field of astrophysical research.  Specifically, our investigation focuses on evaluating the distinguishability among models concerning their equation of state, progenitor mass, and the considered mixing scheme. We found that HK stands out as the optimal detector for distinguishing between models in these three categories, making it the most relevant detector for model differentiation. Conversely, JUNO and DUNE exhibit limited sensitivity in distinguishing between equations of state and progenitor masses. However, they demonstrate significant capability in discriminating between mass ordering, whether through an analysis of the complete signal or by focusing exclusively on the neutronization burst phase. If all three detectors operate simultaneously in a complementary fashion during a future SN explosion, we will significantly enhance our ability to distinguish about this various aspects. This heightened capacity for differentiation stems from the collective power and diverse perspectives provided by the synchronized operation of these detectors. 

Our current study has the potential for expansion by replicating its methodology with a larger and more diverse sample, encompassing a broader array of models. Additionally, a valuable augmentation to our research would involve investigating the systematic errors inherent in the observed correlations. We are optimistic about addressing these aspects in forthcoming endeavors, as this current work serves as an initial stride in that trajectory.
Furthermore, we leave for a future investigation the examination of collective effects, background noises, and longer times simulations. Specifically, we aim to delve into the study of rapid flavor transformations and their potential impact on the expected signal.


\begin{acknowledgments}
Acknowledgments: We thank to the Interdisciplinary Theoretical and Mathematical Sciences Program (iTHEMS) at RIKEN, the NSF Network for Neutrinos, Nuclear Astrophysics, and Symmetries (N3AS) Physics Frontier Center, and the RIKEN Astrophysical Big Bang Laboratory for providing spaces for enriching discussions during the development of this research. We also thank Hiroki Nagakura for the fruitful discussions.
A.H. was supported by JSPS KAKENHI Grant Numbers JP20H01905 and JP21K13913.
S.N. was supported by JSPS KAKENHI (A) Grant Number JP19H00693 and RIKEN Pioneering Project for Evolution of Matter in the Universe (r-EMU).
We acknowledge the high-performance computing resources of the K-computer / the supercomputer Fugaku provided by RIKEN, the FX10 provided by Tokyo University, the FX100 provided by Nagoya University, the Grand Chariot provided by Hokkaido University, and  the FUJITSU Supercomputer PRIMEHPC FX1000 (Wisteria/BDEC-01)/Oakforest-PACS provided by JCAHPC through the HPCI System Research Project (Project ID: hp130025, 140211, 150225, 150262, 160071, 160211, 170031, 170230, 170304, 180111, 180179, 180239, 190100, 190160, 200102, 200124, 210050, 210164, 220047, 220173, 220223, 230056, 230204, 230270) for producing the 2D model dataset.
\end{acknowledgments}

%

\vspace{5mm}



\appendix

\section{Supplementary data on total counts and neutrino-energy relations} \label{appB}

The table \ref{table:events_real} shows the expected number of total events associated with all the studied models, for the case of realistic detectors.  

\begin{table}[ht!]
\begin{center}
\caption{Total number of events expected in each studied channel considering realistic detectors for all the studied models. The counts are given for the cases without considering oscillations (NO), normal mass ordering (NMO) and inverted mass ordering (IMO).}
\vspace{0.1cm}
{\renewcommand{\arraystretch}{0.90}
{\fontsize{8}{8}\selectfont \begin{tabular}{ccccc}
\hline
SN model & $\braket{N}_{\rm HK-IBD}$& $\braket{N}_{\rm DUNE-{\nu_e\ +  Ar}}$ & $\braket{N}_{\rm JUNO-IBD}$ & $\braket{N}_{\rm JUNO-pES}$  \\ \hline
FS 11.2M$_\odot$ & & & &   \\
NO & 4060.73  & 263.57 & 300.86 & 94.52  \\
NMO &4135.24  &  232.35 &  321.68 & \\
IMO & 4287.04  &241.24  & 364.11  &  \\
FT 11.2M$_\odot$ & & & &   \\
NO & 4412.90 & 262.40 & 328.97 & 103.67\\
NMO & 4399.86 & 227.22 & 343.80& \\
IMO &  4373.35 & 237.25  & 374.0 &   \\
FT 15M$_\odot$ & & & &   \\
NO & 8729.38  &  534.59& 665.66 & 180.81 \\
NMO & 7972.33 & 308.52& 630.12 &   \\
IMO & 6430.36  & 372.89 & 557.75 &   \\
FT 27M$_\odot$ & & & &   \\
NO &  8001.56 & 500.24   &  603.00 &  181.38\\
NMO & 7597.67& 330.55 & 597.60 &  \\
IMO & 6775.03 & 378.87 & 586.65 &   \\
LS 11.2M$_\odot$ & & & &   \\
NO & 3698.63 & 262.58 & 271.05 & 90.56 \\
NMO &3834.93& 220.77 & 296.67&   \\
IMO &  4112.51 & 232.63&  348.84 &   \\
LS 15M$_\odot$ & & & &   \\
NO & 7205.00 & 509.16  & 535.97 & 150.0\\
NMO &6743.71 &290.45 &523.79  &  \\
IMO &5804.10 &352.57&498.98 &   \\
LS 27M$_\odot$ & & & &   \\
NO &  6865.32  &  499.00 & 507.86 & 155.38\\
NMO &6633.72  & 311.00 &  514.81&  \\
IMO &6162.03 & 364.19& 528.99 &   \\
FS 11.2M$_\odot$ rot & & & &   \\
NO & 2682.33 & 174.32 & 194.45& 62.65 \\
NMO & 2783.44 & 164.33 &213.83   & \\
IMO &  2989.51  & 167.17   & 253.31 &   \\
FT 15M$_\odot$ rot & & & &   \\
NO & 6156.31 &344.28 & 452.64& 124.11 \\
NMO & 5780.72 & 251.82 & 445.79 & \\
IMO & 5015.67 & 278.14& 431.83 &   \\
3D 9M$_\odot$ & & & &   \\
NO & 5803.79 & 319.65 & 448.43 & 126.17 \\
NMO & 5526.81& 256.74 & 441.38  &\\
IMO &5013 & 263.57 & 300.86  &   \\
3D 10M$_\odot$ & & & &   \\
NO & 7349.60 & 434.14   & 579.05  & 162.63  \\
NMO & 6793.82 & 278.54 & 550.26  &  \\
IMO & 5661.89  & 322.84  &  328.97&   \\
3D 12M$_\odot$ & & & &   \\
NO & 7351.13 &430.19 & 576.43 & 160.64  \\
NMO & 6824.39&286.21 &550.70 &  \\
IMO &5751.68 &  534.59  &  665.66 & \\
3D 13M$_\odot$ & & & &   \\
NO &  9916.66 &  593.40 &798.33 & 236.19\\
NMO &8944.62 & 302.16  & 737.07&  \\
IMO & 6965.02& 385.08& 612.31 &  \\
3D 14M$_\odot$ & & & &   \\
NO &10690.80 &624.14  & 870.83 & 266.51\\
NMO &9525.98  &  281.87& 792.17&  \\
IMO & 7153.43 & 379.32&  632.21&  \\
3D 15M$_\odot$ & & & &   \\
NO & 8917.34  & 528.16 & 713.04 & 202.60  \\
NMO &  8054.11&288.13 & 659.56&   \\
IMO & 6296.01 & 356.47  & 550.63 &   \\
3D 19M$_\odot$ & & & &   \\
NO & 9747.49 & 589.19 & 777.31  & 228.42  \\
NMO &8920.65  & 331.84&  730.03 &   \\
IMO &7236.70&  405.11 & 633.74  &   \\
3D 25M$_\odot$ & & & &   \\
NO & 13692.27 &  811.58& 1125.61&  371.30\\
NMO & 12235.23 & 329.90 & 1026.43 &   \\
IMO & 9267.60  & 467.04  & 825.06 &   \\
3D 60M$_\odot$ & & & &   \\
NO & 8896.73   & 532.58 & 701.21 & 202.49  \\
NMO &8242.20 &232.35  & 668.41  \\
IMO & 6909.17 &  389.45&  601.60 &   \\
\hline
\end{tabular}}\label{table:events_real}}
\end{center}
\end{table}

\begin{figure}[h]
  \centering
        \includegraphics[width=.4\textwidth]{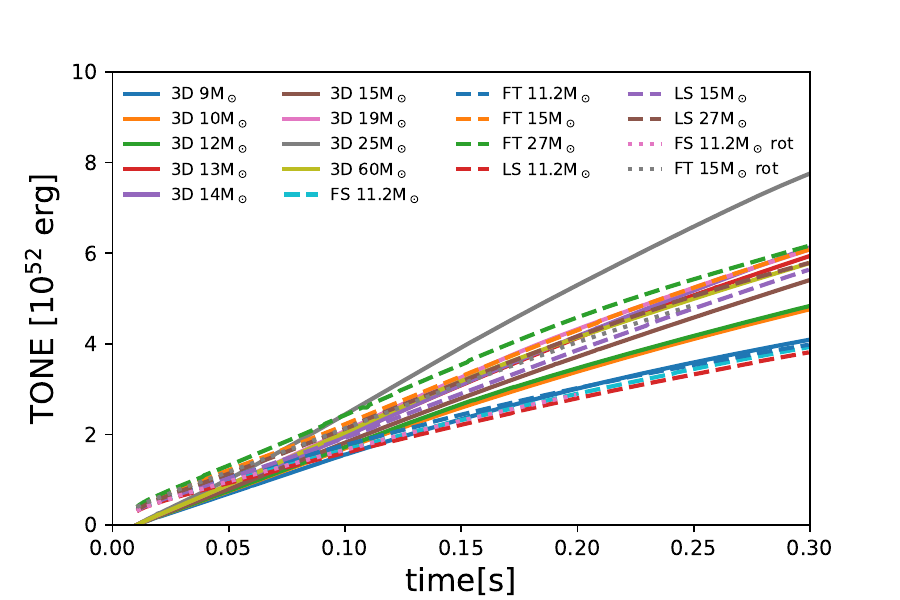} \caption{TONE as a function of time for all the models of our sample.}\label{timevstone}
    \end{figure}

As mentioned earlier, TONE is a time-dependent quantity, representing the total emitted neutrino energy up to a specific post-bounce time. In Figure \ref{timevstone}, the variation of TONE with time for each of the studied models is illustrated.

In Equations \ref{fits}, we present the derived polynomials describing the correlation between TONE and cumulative counts. These equations are specific to different detectors and mass orderings, aligning with the patterns observed in Figures \ref{tone_counts_real_HK}, \ref{tone_counts_real_DUNE}, and \ref{tone_counts_real_JUNO}. 

\begin{eqnarray}\label{fits}
    \rm{Cum}_{{(HK-NMO)}_{3D}}&=&  (52.5E^2 + 1195 E)\left(\frac{d}{10kpc}\right)^{-2} \left(\frac{V}{188 kt}\right) \nonumber \\
        \rm{Cum}_{{(HK-IMO)}_{3D}}&=&  (1198E - 171.77) \left(\frac{d}{10kpc}\right)^{-2} \left(\frac{V}{188 kt}\right)\nonumber \\
            \rm{Cum}_{{(HK-NMO)}_{2D}}&=&  (97.34E^2 + 680.27 E) \left(\frac{d}{10kpc}\right)^{-2} \left(\frac{V}{188 kt}\right)\nonumber \\
        \rm{Cum}_{{(HK-IMO)}_{2D}}&=& (1143.7E - 506.44)\left(\frac{d}{10kpc}\right)^{-2} \left(\frac{V}{188 kt}\right)\nonumber \\
            \rm{Cum}_{(DUNE-NMO)}&=&  (0.43E^2 + 51.19 E)\left(\frac{d}{10kpc}\right)^{-2} \left(\frac{V}{40 kt}\right)\nonumber \\ 
        \rm{Cum}_{(DUNE-IMO)}&=&  (59.63E + 9.91)\left(\frac{d}{10kpc}\right)^{-2} \left(\frac{V}{40 kt}\right)\nonumber \\
      \rm{Cum}_{{(JUNO-IBD-NMO)}_{3D}}&=&  (5.38E^2 + 92.6 E)\left(\frac{d}{10kpc}\right)^{-2} \left(\frac{V}{18.25 kt}\right)\nonumber \\ 
      \rm{Cum}_{{(JUNO-IBD-IMO)}_{3D}}&=&  (106.8E - 19.37)\left(\frac{d}{10kpc}\right)^{-2} \left(\frac{V}{18.25 kt}\right)\nonumber \\ 
      \rm{Cum}_{{(JUNO-IBD-NMO)}_{2D}}&=&  (8.05E^2 + 50.75 E)\left(\frac{d}{10kpc}\right)^{-2} \left(\frac{V}{18.25 kt}\right)\nonumber \\ 
      \rm{Cum}_{{(JUNO-IBD-IMO)}_{2D}}&=&  (99.26E - 45.5)\left(\frac{d}{10kpc}\right)^{-2} \left(\frac{V}{18.25 kt}\right)\nonumber \\ 
      \rm{Cum}_{{(JUNO-pES)}_{3D}}&=&  (2.31E^2 + 26.12 E)\left(\frac{d}{10kpc}\right)^{-2} \left(\frac{V}{18.25 kt}\right)\nonumber \\ 
      \rm{Cum}_{{(JUNO-pES)}_{2D}}&=&  (1.93E^2 + 17.83 E )\left(\frac{d}{10kpc}\right)^{-2} \left(\frac{V}{18.25 kt}\right)
\end{eqnarray}

\section{Bayes factor calculation: example case} \label{appA}
Illustrating our Bayesian methodology, we delve into the subsequent example. In this case, we undertake a comparison between the 2D models FS 11.2 M$_\odot$ and FT 11.2 M$_\odot$, specifically focusing on the IBD channel at HK. This case corresponds to the first entry within Table \ref{table:ln_bayes_factor_EoS}.

Initially, we posit the FS 11.2 M$_\odot$ model as our reference and initiate a comparison with its counterpart. Subsequently, we interchange the roles, leading to an asymmetric outcome, as expected.

Following the exposition in Section \ref{sec:model_discr}, we proceed to evaluate our ability to discern the authentic model from alternative ones. To this end, we employ a Monte Carlo simulation to generate a sample comprising $10^4$ instances of the signal within our detector. Subsequently, we calculate both the mean ($\langle \ln\mathcal{B}_{{\alpha\beta}} \rangle$) and the standard deviation ($\sigma (\ln\mathcal{B} {_{\alpha\beta}})$) for scenarios involving the same two models, while interchanging their roles as the true model. Table \ref{table:example_case}  summarizes the outcomes for these two distinct cases.

\begin{figure}[h]
  \centering
        \includegraphics[width=.5\textwidth]{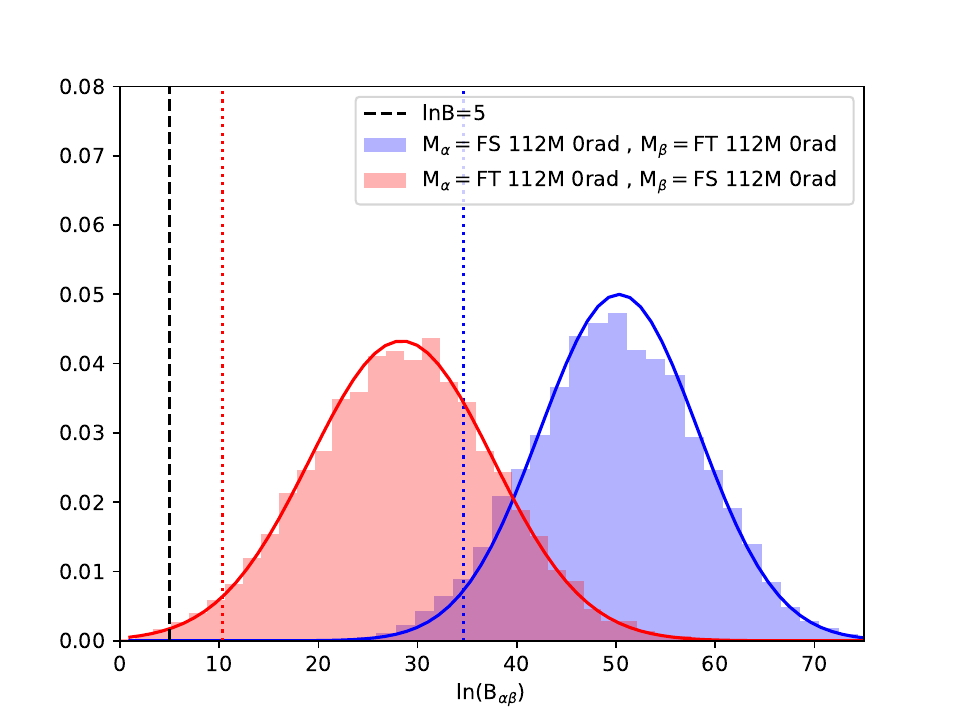} \caption{Histograms of $\ln(\mathcal{B}_{\alpha\beta})$ distributions for the two pairs associated with models: FS 11.2 M$_\odot$ and FT 11.2 M$_\odot$, considering  the HK detector and the no-oscillation scenario.  }\label{histograms}
    \end{figure}

Figure \ref{histograms} visually represents the binned outcomes of the Monte Carlo sample for both feasible model pairs. Notably, these distributions are well aligned with the normal distribution curve, characterized by corresponding values of $\langle \ln\mathcal{B}{_{\alpha\beta}} \rangle$ and $\sigma (\ln\mathcal{B} {_{\alpha\beta}})$. The dotted lines on the graph indicate $\langle \ln\mathcal{B}{{_{\alpha\beta}}} \rangle - 1.96\sigma (\ln\mathcal{B}_{{\alpha\beta}}) $ for both distributions.

\begin{table}[ht!]
\begin{center}
\caption{The mean Bayes factors along with standard deviations calculated with M$_\alpha$ being the true model, using $10^4$ simulated signals in the IBD channel at HK detector from an SN at $d = 10$ kpc. The true model can be distinguished from the alternative at the  95\% CL
for $\langle \ln\mathcal{B}_{{\alpha\beta}} \rangle - 1.96\sigma (\ln\mathcal{B}_{{\alpha\beta}})  > 5$ , which is satisfied by all the
entries.}
\vspace{0.5cm}
{\renewcommand{\arraystretch}{1}
\begin{tabular}{ccc}
\hline
M$_\alpha$ &  M$_\beta$  & $\langle \ln\mathcal{B}_{{\alpha\beta}} \rangle \pm \sigma (\ln\mathcal{B} {_{\alpha\beta}})$\\ \hline
 FS 11.2M$_\odot$ & FT 11.2M$_\odot$  &  50.31 $\pm$  7.98\\
FT 11.2M$_\odot$ & FS 11.2M$_\odot$  & 28.38 $\pm$ 9.22 \\
\hline
\end{tabular}}\label{table:example_case}
\end{center}
\end{table}

Finally, in order to symmetrize the results and facilitate their interpretation, we combine these two cases by calculating their average\footnote{The complete tables prior to the symmetrization process can be requested from the corresponding author}. This process yields the conclusive $\langle \overline{\ln\mathcal{B}{_{\alpha,\beta}}} \rangle$ and $\sigma (\overline{\ln\mathcal{B} {_{\alpha,\beta}}})$, which serve as the foundation for assessing the distinguishability between the FS 11.2 M$_\odot$ and FT 11.2 M$_\odot$ models. In this specific instance, the computed values are $\langle \overline{\ln\mathcal{B}{_{\alpha,\beta}}} \rangle = 39.34$ and $\sigma (\overline{\ln\mathcal{B} {_{\alpha,\beta}}}) = 8.62$ (first entry of Table \ref{table:ln_bayes_factor_EoS}), which meet our predefined criteria.

\newpage

\bibliography{main}{}
\bibliographystyle{aasjournal}



\end{document}